\documentclass[trackchanges]{aastex63}
\usepackage{xcolor}
\usepackage{threeparttable}
\usepackage{amsmath}

\accepted{January 28, 2022}
\submitjournal{ApJ}

\shorttitle{Low Energy Flares in the GJ 65 System}
\shortauthors{Fleming et al.}

\begin{document}

\title{New Time-Resolved, Multi-Band Flares In The GJ 65 System With gPhoton}

\correspondingauthor{Scott W. Fleming}
\email{fleming@stsci.edu}

\author[0000-0003-0556-027X]{Scott W. Fleming}
\affiliation{Space Telescope Science Institute, 3700 San Martin Dr, Baltimore, MD 21218, USA}

\author[0000-0003-2732-3486]{Chase Million}
\affil{Million Concepts LLC, 1355 Bardstown Rd., No. 132, Louisville KY 40204}

\author[0000-0001-5643-8421]{Rachel A. Osten}
\affiliation{Space Telescope Science Institute, 3700 San Martin Dr, Baltimore, MD 21218, USA}
\affiliation{Center for Astrophysical Sciences, Department of Physics \& Astronomy, Johns Hopkins University, 3400 North Charles St., Baltimore, MD 21218, USA}

\author[0000-0002-0687-6172]{Dmitrii Y. Kolotkov}
\affiliation{Centre for Fusion, Space and Astrophysics, Department of Physics, University of Warwick, Coventry CV4 7AL, UK}
\affiliation{Institute of Solar-Terrestrial Physics, Lermontov St., 126a, Irkutsk 664033, Russia}

\author[0000-0002-9314-960X]{C. E. Brasseur}
\affiliation{Space Telescope Science Institute, 3700 San Martin Dr, Baltimore, MD 21218, USA}
\affiliation{SUPA Physics and Astronomy, University of St. Andrews, Fife, KY16 9SS, Scotland, UK}



\begin{abstract}
Characterizing the distribution of flare properties and occurrence rates is important for understanding habitability of M dwarf exoplanets. The GALEX space telescope observed the GJ 65 system, composed of the active, flaring M stars BL Cet and UV Cet, for 15900 seconds ($\sim$4.4 hours) in two ultraviolet bands. The contrast in flux between flares and the photospheres of cool stars is maximized at ultraviolet wavelengths, and GJ 65 is the brightest and nearest flaring M dwarf system with significant GALEX coverage. It therefore represents the best opportunity to measure low energy flares with GALEX. We construct high cadence light curves from calibrated photon events and find 13 new flare events with NUV energies ranging from $10^{28.5} - 10^{29.5}$ ergs and recover one previously reported flare with an energy of $10^{31}$ ergs. The newly reported flares are among the smallest M dwarf flares observed in the ultraviolet with sufficient time resolution to discern light curve morphology. The estimated flare frequency at these low energies is consistent with extrapolation from the distributions of higher-energy flares on active M dwarfs measured by other surveys. The largest flare in our sample is bright enough to exceed the local non-linearity threshold of the GALEX detectors, which precludes color analysis. However, we detect quasi-periodic pulsations (QPP) during this flare in both the FUV and NUV bands at a period of $\sim$50 seconds, which we interpret as a modulation of the flare's chromospheric thermal emission through periodic triggering of reconnection by external MHD oscillations in the corona. 
\end{abstract}

\keywords{stars: flare -- stars: individual (GJ65 AB (BL Cet, UV Cet)) -- stars: late-type}


\section{Introduction} \label{sec:intro}
\subsection{Stellar Flares in the Ultraviolet}
Stellar flares may have a substantial effect on the habitability of planets that orbit them closely. Although a single stellar flare is unlikely to have a lasting impact on a planet's atmosphere \citep{2010AsBio..10..751S}, the cumulative effects from repeated flares can drive atmospheric chemistry \citep[e.g.,][]{2007AsBio...7...30T,2016ApJ...830...77V,2017A&A...608A..75C,2017ARA&A..55..433K,2018ApJ...867...70L,2019ApJ...871L..26F}, cause atmospheric loss \citep[e.g.,][]{2016A&A...596A.111R,2018PNAS..115..260D,2019AsBio..19...64T}, and could hinder \citep{2007AsBio...7..185L,2017MNRAS.465L..34A,2017MNRAS.469L..26O,2019ApJ...881..114Y} or enhance \citep{2016AsBio..16...68R,2017ApJ...843..110R} biogenesis. Flare rates on the Sun follow a power-law as a function of flare energy \citep{2011SSRv..159..263H,2012JGRA..117.8103S}, such that more energetic flares happen less often than smaller energy flares. Our understanding of the flare frequency distribution (FFD) is rapidly improving for G, K, and M dwarfs thanks to both ground-based \citep{2011PhDT.......144H,2012ApJ...748...58D,2019ApJ...881....9H,2020MNRAS.491...39C} and space-based missions like Kepler/K2 \citep{2014ApJ...797..121H,2015EP&S...67...59M,2016ApJ...829...23D,2017ApJ...845...33G,2017ApJ...849...36Y,2020A&A...637A..22R}, HST \citep{2012ApJ...754....4O,2018ApJ...867...71L,2018ApJ...867...70L}, and TESS \citep{2019MNRAS.tmp.2115D,2020AJ....159...60G}. Even with the long baselines and minute-cadence observations afforded by space telescopes, most detected flares and flare rate estimates from these large UV and optical surveys have been at (non-bolometric) energies $\log{E} \gtrsim 30$. Flare energy tends to scale directly with both peak brightness and duration in the optical  \citep{2015EP&S...67...59M,2017ApJ...851...91N}, so the ability of space-based surveys like Kepler and TESS to detect low energy flares is limited by relative magnitude sensitivity and cadence. Indeed, the cadence of observations can impact the measurement accuracy of even large energy flares.  \citet{2018ApJ...859...87Y} found that Kepler long-cadence data sampled at 30 minutes compared to short-cadence data sampled at one minute resulted in energies underestimated by 25\% and systematically underestimated amplitudes and overestimated durations by 50\%.  \citet{2020A&A...637A..22R} compared short- and long-cadence K2 light curves of flares on M dwarfs, and found that flare amplitudes were underestimated by 30\% and flare durations were overestimated by 60\%.  However, these effects offset one another when deriving their flare energies, which used an equivalent duration (area under the curve).  Flares in the ultraviolet (UV) may not follow as direct a relationship seen in optical surveys, where the flares in UV may appear shorter in duration and not tied as closely to energy across comparable integrated energy regimes \citep[see Figure 21 in][]{brasseur2019}.

Observations in the far-ultraviolet (FUV) and near-ultraviolet (NUV) provide unique constraints to flare physics and any interactions they might have with planetary atmospheres compared to other wavelength regimes.  The UV fluxes originate from the upper atmosphere of the star, namely, the chromosphere and transition region leading to the corona.  In the FUV, stellar fluxes can be dominated by Lyman alpha emission \citep{2013ApJ...766...69L}, which can photodissociate $\mathrm{H_{2}O}$, $\mathrm{CH_4}$, and $\mathrm{CO_2}$ in exoplanet atmospheres \citep{2006PNAS..10318035T}.  Other major lines come from C IV, Si IV, He II, and Al II \citep{2008ApJS..175..229A,2009A&A...498..915D,2016ApJ...830..154F}.  In the NUV, the primary contributions come from the Mg II h and k doublet, along with Al and ionized Fe lines \citep{2019ApJ...871..235P}, although the GALEX NUV band has low transmission near the Mg II lines, and thus is more likely to be dominated by continuum emission \citep{2005ApJ...633..447R}.  While these line emissions can be present when a flare is not happening, the line emission increases during a flare due to the additional hot plasma generated by a flare.  Furthermore, a hot blackbody component from the flare event contributes to the overall emission, which is not present before or after the flare event.  Observations across a wide range of wavelengths, including X-ray, UV, optical, and radio probe different physics during the flare events \citep{1991ARA&A..29..275H,1996A&A...310..908V,2012SoPh..277...21K,2013ApJS..207...15K}, especially if simultaneous observations of a flare can be obtained.  Previous studies have used photon events from GALEX to search for intra-visit variable sources \citep{2005AJ....130..825W}, especially to detect and characterize flares on individual stars \citep{2005ApJ...633..447R, 2006A&A...458..921W}, and one survey detected 52 flare events on 49 stars using 1802 NUV images \citep{2007ApJS..173..673W}. \citet{2018MNRAS.475.2842D} reported detections of quasi-periodic pulsations (QPP) in flares from six M dwarfs, including all four of the M dwarfs that had flares reported in \citet{2006A&A...458..921W}.

\subsection{The GJ 65 System}
The GJ 65 system is composed of two M dwarfs with an on-sky separation of 2.26$\arcsec$ \citep{2018A&A...616A...1G}, unresolved by the 4.2$\arcsec$ and 5.3$\arcsec$ full-width half maxima image resolutions of GALEX in the FUV and NUV bands \citep{2007ApJS..173..682M}.  The two components have Gaia-based distance estimates of $2.687 \pm 0.002$ and $2.703 \pm 0.002$ pc \citep{2018AJ....156...58B}.  They orbit in an eccentric ($e \sim$0.6) 26.3-year orbit and have a combined mass of $\sim 0.24 \; \rm{M_{\odot}}$ \citep{2019ApJ...871...63M}.  \citet{2017MNRAS.471..811B} used Doppler image reconstructions to measure the rotation rates of the two components (0.24 and 0.23 days, respectively).  \citet{2016A&A...593A.127K} were able to measure the angular diameters of the two components using interferometry, deriving radii of 0.165 and 0.159 $R_{\odot}$ to 4\% relative precision.  Using adaptive optics, the same authors derived masses for the two components of 0.1225 and 0.1195 $M_{\odot}$, also at 4\% relative precision.  Given these masses, the radii differ from model predictions at the 10-15\% level, a common observation for low-mass stars with strong magnetic fields and stellar activity.  \citet{2017ApJ...835L...4K} used spectro-polarimetric observations to study the magnetic topology of the two components.  Despite having very similar masses and radii, BL Cet's total magnetic energy is an order of magnitude smaller than UV Cet, and while BL Cet exhibits a complex field structure, UV Cet has a strong, axisymmetric dipolar magnetic field.

The GJ 65 stars are among the closest, brightest, active flare stars with a large amount of GALEX coverage\footnote{The only other active M dwarfs within 5 pc that have multiple hours of GALEX coverage is the Wolf 424 system, which has $\sim$7500 seconds of coverage and a similar GALEX NUV magnitude. Like the GJ 65 system, it is also composed of two M dwarfs, and they orbit with a comparable period to GJ 65 ($\sim$15.5 years).  We find six flares with energies between $28.8 < \log(E) < 30.1$, to be described in a future article.}, and thus presents the best opportunity to search for and study time-resolved, short-duration stellar flares in the GALEX UV bands. High-cadence searches in the optical, with time resolutions of seconds or even microseconds, have been conducted on the GJ 65 system \citep{1973ApJ...185..239B,2016A&A...589A..48S,2017PASA...34...10B}.  The survey by \citet{2007ApJS..173..673W} did not include GJ 65 system. From the visit level photometry catalogs, \citet{2017AJ....154...67M} detected a flare in the GJ 65 system, but did not analyze any photon event-level data.  \citet{2018MNRAS.475.2842D} analyzed GJ 65, in the near-UV only, with the gPhoton software using one-second time resolution bins, detecting a QPP in the largest flare via a wavelet analysis.  In this paper, we present results from a time-resolved search for GJ 65 flares using all available GALEX archival photometry.

In Section \ref{sec:style} we describe the data products and processing. In Section \ref{sec:detections} we describe our algorithms for detecting flare events within the GALEX light curves. In Section \ref{sec:flaredesc} we present the detected flare events and further analyze the relatively large flare previously found by \citet{2017AJ....154...67M}, including a discussion on how the brightness of the large flare precludes an analysis of the FUV-NUV color of the flare, and our detection and anaysis of the QPP during the flare.

\section{Observations and Data Processing} \label{sec:style}
GALEX \citep{2005ApJ...619L...1M, 2007ApJS..173..682M} collected data with photon-counting micro-channel plate detectors in two ultraviolet bands. The Far Ultraviolet (FUV) band extended from $\sim$1344 \text{\AA} to $\sim$1786 \text{\AA}, and the Near Ultraviolet (NUV) band extended from $\sim$1771 \text{\AA} to $\sim$2831 \text{\AA}.  Observations were organized into ``visits'': one or more sequences of observations conducted while the spacecraft was behind Earth's shadow, lasting in total no more than approximately 30 minutes. GALEX had both imaging and spectroscopic capabilities, and the imaging data were generally released to the research community as integrated images with detected source catalogs. The mission occasionally provided photon-level data to researchers by special request. We use data available through the open-source gPhoton project \citep{2016ApJ...833..292M}, which re-created portions of the GALEX mission pipeline to generate and archive calibrated photon events from all observations taken through GALEX Data Release 7.

The present work analyzes dual-band observations of the GJ 65 system across ten visits. Nine of these were $\sim$30 minute observations made within a span of two days (18-19 November 2005). The tenth visit is a guest investigator (GI) observation made on 09 October 2016 with an integration time of about 10 minutes.  Two additional observations were made in single-band ``AIS'' observing mode and have integration times of less than two minutes each. These visits are too short for a meaningful flare measurement, even if a flare occurred. We visually inspected light curves for these shorter visits, verified that no obvious flare behavior was present, and then eliminated them from further analysis. Removing these visits from the sample reduced the total effective observation time by $\sim$5\%.  The total observation time for the ten visits\footnote{Although gPhoton uses the uncalibrated photon events and the aspect solution files to create the images and light curves analyzed here, the mission-produced pipeline products corresponding to the visits we examined are available at MAST via \doi{10.17909/t9-638a-q564}.} used in this analysis (ignoring the two AIS visits) is 15682 seconds ($\sim$4.36 hours).

\subsection{Data Processing}
\subsubsection{Photometric Measurements}
Our photometry is derived from photon events processed using a special branch of gPhoton, ``v1.28.9\_nomask''. Due to our special treatment of hotspot masks (see Section \ref{hotspotsec}), we start from the raw photon events themselves, perform calibration of the photon events, and extract light curves without using the standard \texttt{gAperture} module.  This branch of gPhoton is setup to allow such analysis. Python notebooks\footnote{Notebook GitHub repository:\url{https://github.com/MillionConcepts/gfcat_gj65} DOI:  \dataset[10.5281/zenodo.3870757]{\doi{10.5281/zenodo.3870757}}} are provided as supplemental material that can be used to re-create our analysis (data products, the flare table, and each figure) using this fork of the gPhoton software. The light curves are extracted using an aperture radius of $17.3 \arcsec$ (equivalent to ``APER7'' in the GALEX catalogs) at a time bin of 30 seconds, except where noted.  The bin size of 30 seconds is chosen such that short-duration flares lasting only a few minutes will have more than one flux bin during the event, while also providing sufficient counts in each time bin to get reasonable signal-to-noise given the brightness of the GJ 65 system. At the NUV magnitude of GJ 65 of $\sim$18, this bin size yields a 3-$\sigma$ error of $\sim$0.25 mag \citep[see Fig.\ 11 from][]{2016ApJ...833..292M}. We did not include a background measurement in these calculations because the quantity of interest for flare-detection is the change in magnitude (which automatically accounts for a stationary sky) and the background flux is a very minor contribution, approximately 27.5 AB Mag in NUV and 29 AB Mag in FUV at the location of GJ 65 from the GALEX mission catalogs.  Note that gPhoton outputs flux densities measured in $\mathrm{erg \; s^{-1} \; cm^{-2} \; \text{\AA}^{-1}}$.

The initial photometric aperture location is based on the GAIA \citep{2018A&A...616A...1G} position, which we then adjust to better match the position in the GALEX observations which were taken 10 years earlier than the 2015.5 epoch that Gaia coordinates are listed in. This initial photometric aperture position is refined once for each visit by re-centering on the median event position within the aperture. The re-centering corrects for the small amount of scatter in GALEX astrometry between visits, and would also correct for proper motion between visits if any were present \citep{2007ApJS..173..682M}.  This re-centering results in a change of the photometric aperture center between 2.5 and 4 arcseconds from visit to visit, well within the GALEX spatial resolution. \citet{2016ApJ...833..292M} caution that pixel-to-pixel variability in the instrument flat can be a significant source of additional uncertainty in sub-visit light curves, and thus can produce false variable or periodic signals. We compare each visit's flux density as a function of time with other output parameters from gPhoton to check for any correlated variability, including the effective exposure times per bin and distance from center of the field.

\subsubsection{Turning Off the Hotspot Mask}
\label{hotspotsec}
Photon events that fall on regions of the detector covered by the GALEX hotspot masks are excluded from calibration and reduction in gPhoton by default. Therefore, the regions of the detector that are masked have an effective response of exactly zero. This approach to hotspot masking generally has a negligible effect on data quality in the mission-produced images and catalogs because the contribution of the masked region was diluted over the duration of an observation by the relative motion produced by spacecraft dither and then corrected by a relative response map that accounted for the effect of the mask. This does not work for shorter integration times; if the light from a source intersects a masked detector region, its flux density will be incorrectly measured. In sub-visit light curves, the interference of a hotspot mask can produce significant non-astrophysical variability over the period of the GALEX $\sim$120 second dither pattern \citep{2017ApJ...845..171B,2018ApJS..238...25D}. The gPhoton software flags any light curve time bin where the photometric aperture contains any photon events falling within one pixel ($\sim$6\arcsec) of the flat field map of a masked region, and it is recommended that researchers exercise additional skepticism when using any flagged data.

\begin{figure}
  \centering
    \includegraphics[width=0.98\textwidth]{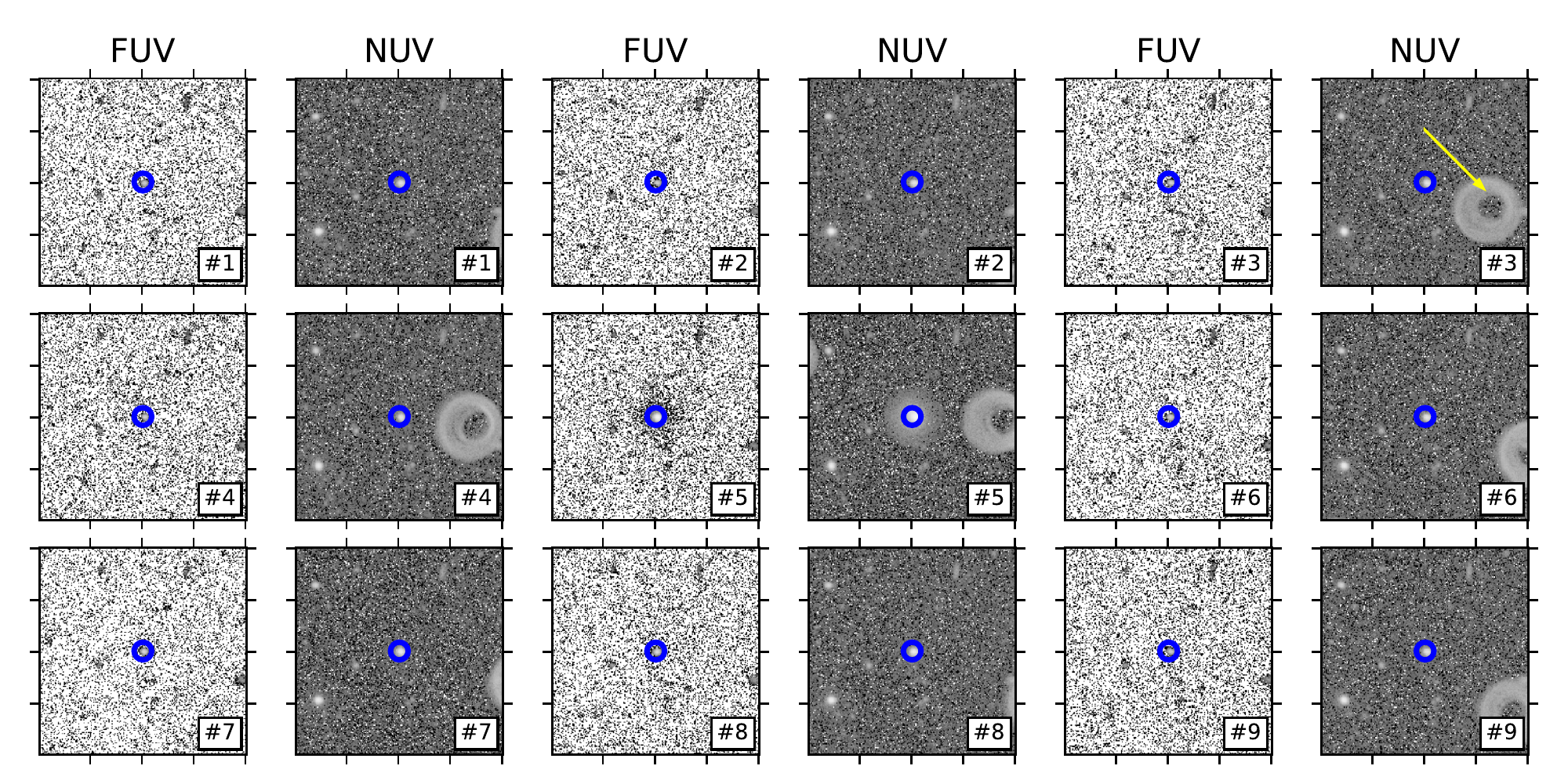}
    \caption{Thumbnail images (FUV, left; NUV, right) of the first nine visits, centered on GJ 65.  Blue circles represent the photometric aperture of 17.3 arcseconds, equivalent to the GALEX MCAT ''APER7''.  The blurred doughnut shapes to the right are hotspots (NUV Visit 3 is first visit when the hotspot is most visible, denoted with a yellow arrow).  The effect of the largest flare on GJ 65's brightness is readily apparent in the NUV during Visit 5 (center). \label{fig_thumbnails}}
\end{figure}

The majority of the GALEX observations of GJ 65 have a masked hotspot close enough to trigger this flagging.  GALEX hotspots are known to vary over time, which is to say that they may have been active during some periods of the mission and not others. The mission hotspot mask was also optimized to remove any possible spurious signal rather than to conserve usable detector area, and a single hotspot mask was created for each band and applied uniformly over the whole mission without regard to the temporal variation of hotspots. The hotspot mask also only has one quarter of the spatial resolution of the final GALEX imaging products, so there may be cases in which a hotspot passes closely enough to a source of interest to potentially affect a measurement even though any non-astrophysical signal from the hotspot would not leak into a reasonably sized photometric aperture. This was the case for the observations of GJ 65, so we modified the \texttt{PhotonPipe} module of the gPhoton data reduction pipeline to aspect-correct photon events even if they are covered by the hotspot mask, re-reduced the photon-level data from the raw mission files, and then ported functions from the the gPhoton tools as needed to create images and calibrated light curves from these files.  This version of the gPhoton software is available in the ``v1.28.9\_nomask'' branch on GitHub. Figure \ref{fig_thumbnails} demonstrates that for all visits of GJ 65 included in our analysis, accurate photometry is still possible even with a nearby hotspot present. The active hotspot appears in the images as annuli that trace the spiral dither pattern. The hotspot near GJ 65 is especially visible in several of the NUV images, and is denoted by the yellow arrow in the NUV Visit 3 panel for reference.  None of the visits analyzed here have the hotspot region pass within the relatively large $17.3\arcsec$ radius photometric apertures, centered on GJ 65 and denoted in blue.  Thus, the default conservative flagging of the photometry measured from locations near, but not within, hotspot mask regions can safely be ignored here, and the photon events from within our aperture (in blue) are not impacted.

\section{Flare Detections}\label{sec:detections}
\subsection{Flare-Finding Algorithm}
A variety of methods have been used for flare detection, and there is not a clear consensus approach. One common strategy is to search for one or more data points in the light curve that increase in flux density at some level of significance, followed by visual inspection and validation by a human researcher \citep[e.g.,][]{2011AJ....141...50W,2016ApJ...829...23D,brasseur2019}. For single-band observations where counting statistics dominate, we prefer a search algorithm that classifies any two consecutive data points that are at least $3\sigma$ above a baseline level, or any one data point at $5\sigma$ above the baseline level, as likely belonging to a flare. The flare time range (start and end of the flare event) is then extended from these detected peaks both forward and backward in time until the data are indistinguishable from the ``instantaneous non-flare flux'' (INFF, see Section \ref{inffsection}). The flare extent can be difficult to determine, especially for smaller flares and those that do not follow a ``fast rise, exponential decay'' (FRED) shape \citep[see, e.g.][]{2014ApJ...797..122D}.  We define the flare range by stepping the extensions backwards and forwards in time until two consecutive points are less than $1\sigma$ above the INFF. This is similar to the method used by \citet{2011AJ....141...50W} and \citet{brasseur2019}. All of the GALEX visits of GJ 65 that we analyze here are dual-band, however, so we have modified our search criteria to trigger on instances where at least two data points in the NUV light curve rise to $3\sigma$ above the local baseline flux density, \emph{or} that there is a simultaneous $3\sigma$ deviation in at least one data point in both the FUV and NUV bands. This has the benefit of identifying a few smaller events that are likely flares (because they occur simultaneously in both bands) that would not be caught by a single-band search. An algorithmic and visual search did not find any flares that are detected in FUV but not in NUV.  This matches expectations and previous results from flare searches in the UV, where the vast majority of flares are detected in both FUV and NUV when multi-band data are available.  Although FUV-only flares are very rare, recent results have shown at least one example of a significant FUV flare (peak flux enhancement $> 14000$) with optical response only increasing by 0.9\% \citep{2021ApJ...911L..25M}.

We examined light curves from other sources in these GALEX visits to see if any other sources exhibited similar flux density outliers within their light curves.  If multiple sources show similar levels of variability, then some of our flare candidates may be caused by unknown systematic errors.  We exclude sources at the outer 20\% of the GALEX field-of-view, since those often suffer from edge effects.  We also exclude sources at or very near to the local non-linearity regime of the GALEX detectors.  We find no other sources that exhibit extremely large flux density outliers, and only a single source (a binary of two similarly-bright components with a separation marginally resolved by GALEX) that have significant variability at the 3-sigma level at all. Thus, the intra-visit variability we see from GJ 65 is likely coming from the stars themselves and is unlikely to be an unknown systematic related to the processing of the photon events or instrument hardware.

The 30 minute maximum duration of GALEX visits is shorter than the duration of many flares with energies of $\log{E} \gtrsim 32$, under reasonable assumptions about flare light curve morphologies. For flares at or above these energies, a single visit in GALEX will capture only a portion of the flare (most likely, the long tail end of the flare), in which case the flux density during the entire visit would likely be monotonically decreasing throughout the entire visit. We do not see any evidence of this during the GALEX observations of GJ 65.

\subsection{Instantaneous Non-Flare Flux}
\label{inffsection}
When searching for flares, the typical practice is to measure the change in brightness due to a flare as offset from some ``quiescent'' flux.  Over long time baselines, it is possible to identify and account for non-flare stellar variations due to other types of stellar activity, rotational spot modulation, or transiting objects. For example, \citet{2020AJ....159...60G} describes such an approach for flare-finding in TESS. For active stars, such as the GJ 65 system, referring to a baseline flux as ``quiescence'' is inaccurate without a long enough baseline to derive a flux value that takes into account intrinsic variability over, e.g., the star's magnetic cycles. For flare detection, the measurement that is actually desired is the contribution of total stellar flux made by all stellar activity except the flare event itself. We refer to this as ``instantaneous non-flare flux'' (INFF); given perfect knowledge about stellar behavior, the INFF would be equivalent to a light curve where all (and only) contributions from flaring have been removed.  Even this would be technically inaccurate, because the pre-flare stellar fluxes may change due to active region configurations related to specific flare events that are not repeatable over time.

The ability to measure INFF is still impacted, in practice, by both the limitations of the data (e.g. how well the bandpass is calibrated, rate of flux sampling, and observation time baseline), and because it is likely that flares can induce other phenomena that affect total luminosity, including sustained increases in flux, QPPs, or additional flares \citep[e.g.,][]{1981Ap&SS..77..347P, 1990A&A...228..513P, 1995MNRAS.277..423P, 1999ApJ...515..746O, 2018MNRAS.475.2842D}. If only the decay phase of a large flare is observed, it might be easily mistaken for some other type of astrophysical variability such as a starspot rotating into view.

Our approach when defining the INFF for a visit is to identify and measure the flux density during specific time windows that ``look like'' quiescence, i.e. have a slope of approximately zero. We have calculated flare energies using these measurements of INFF. Following the method of \citet{2016ApJ...829...23D}, we use a sigma-clipping algorithm to make an initial estimate of the INFF, which is used to conduct a flare search in the manner described above. A new INFF estimate was then made by running sigma-clipping again on light curves from which flaring time periods were excised, and used as input to a final algorithmic search for flaring. Uncertainties on the INFF ``quiescence'' reflect counting statistics on the integrated value of all non-flare data points.

\subsection{Detection Limit}
Because it relies exclusively on the most extreme outlier, our flare detection algorithm imposes an immediate lower limit on the size of any detectable flare. The smallest detectable flare is one whose light curve deviates precisely by $3\sigma$ from the INFF for exactly one time bin. For light curves with a time resolution of 30 seconds, this flare would be reported with a duration of $\sim$120 seconds. This is also the typical duration of ``AIS'' or ``scan mode'' visits, which is why these have very little use for stellar flare studies.  The detection threshold and encapsulated energy will vary as a function of the INFF. For a peak INFF AB magnitude of 17.77 in NUV, the maximum detectable change in magnitude is 0.188 ($\sim$19\%), with an encapsulated energy in NUV of $\log{E} \sim$28.7 erg. On the lower end, an INFF AB magnitude of 18.66 in NUV has a maximum detectable change in magnitude of 0.27 ($\sim$ 27\%), with an encapsulated energy in NUV of $\log{E} \sim$29.02 erg.

\subsection{Energy Calculation}
\begin{figure}
\includegraphics[width=0.8\textwidth]{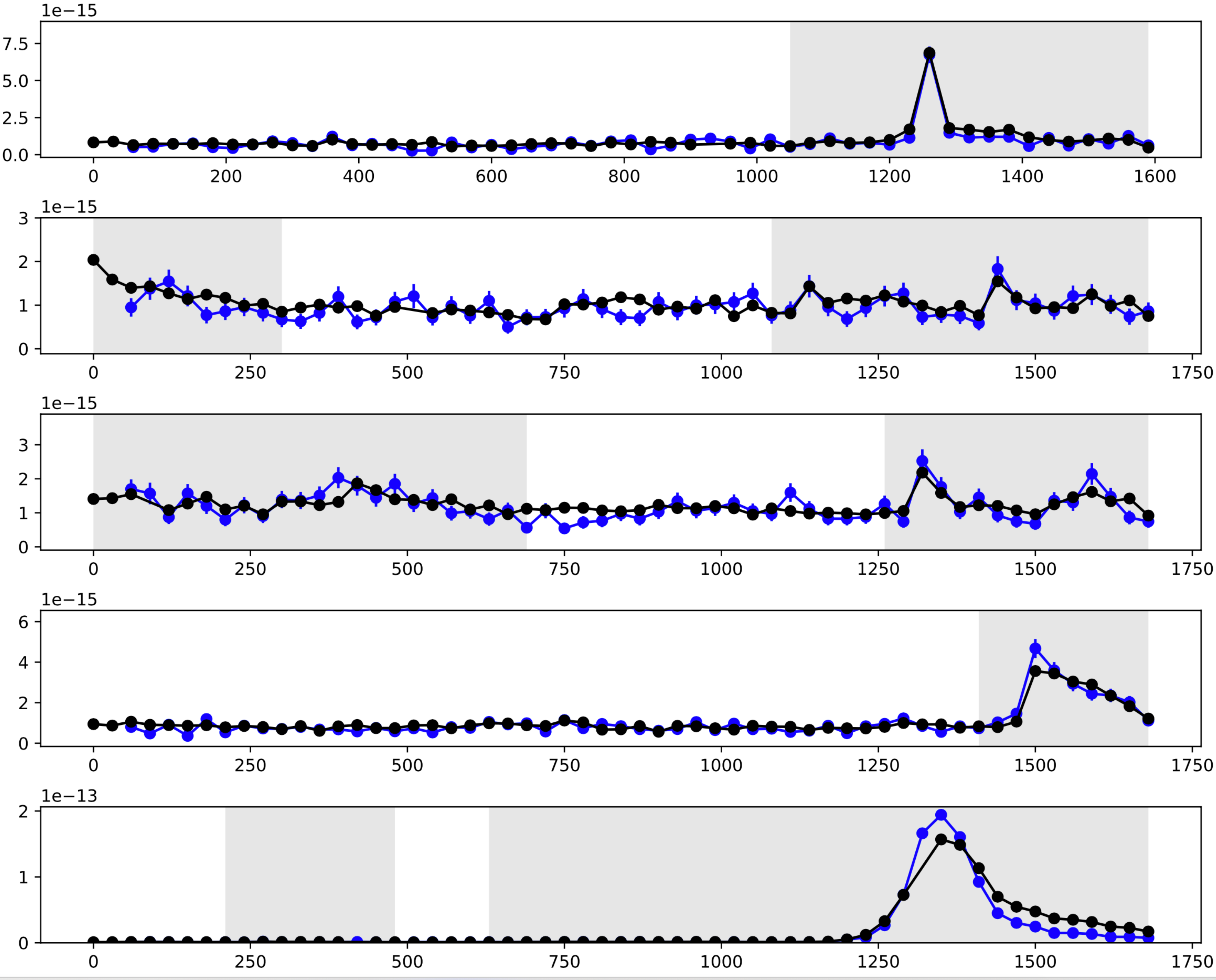}
\caption{Visit-level light curves for Visits 1-5 (top to bottom).  FUV is in blue, NUV is in black, and detected flare events are denoted by the grey backgrounds.  See the Appendix for more detailed plots of each visit and flare.\label{allflares_1}}
\end{figure}

\begin{figure}
\includegraphics[width=0.8\textwidth]{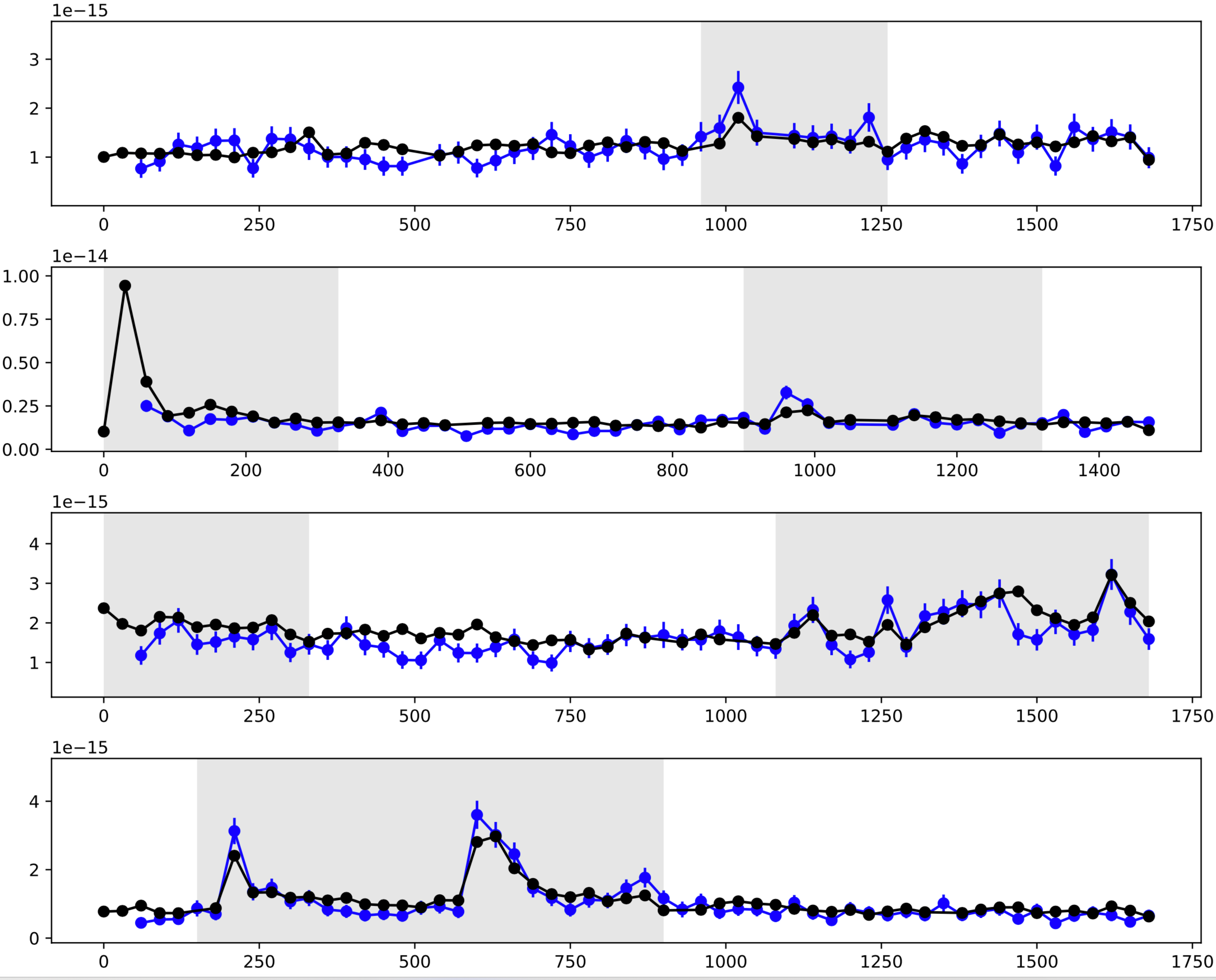}
\caption{Visit-level light curves for Visits 6-9 (top to bottom).  FUV is in blue, NUV is in black, and detected flare events are denoted by the grey backgrounds.  See the Appendix for more detailed plots of each visit and flare.\label{allflares_2}}
\end{figure}

Figures \ref{allflares_1} and \ref{allflares_2} show the FUV (blue) and NUV (black) visit-level light curves, with the identified flare events highlighted in grey. Following \citet{brasseur2019}, we calculate the energies of each flare event by integrating the area under the curve from flare start to flare end as defined by our flare finding algorithm, subtracted by the INFF value for each band.  We convert the integrated flux densities to fluence using GALEX effective widths in FUV and NUV of $255.45 \; \text{\AA}$ and $729.94 \; \text{\AA}$, respectively \citep{svoprof}. We assume a distance of 2.687 pc \citep{2018AJ....156...58B}.  The flare energy measurements are lower limits because of uncertainty in the INFF measurement, but also because it is possible that a small candidate flare event toward the beginning of a visit may actually be a sub-component of a larger event that may have happened before the visit.  More commonly, the entire duration of a candidate flare event is not contained within the visit, and is either truncated at the start or end of the visit.  For some flare events, the flux densities return to nearly the same INFF as before the flare event, and the derived energy is quite close to the total, but in other cases more of the flare's tail is truncated, and thus more of the total energy is missing.

Our energies are reported in the GALEX FUV and NUV bandpasses, and thus would need to be scaled for comparison with energies from other surveys and instruments. To derive an estimate of the fractional GALEX flare energy to bolometric flare energy, we calculate the bolometric conversion factor following \citet{brasseur2019}, resulting in $p_{bol}$ values of 0.021 and 0.133 for FUV and NUV, respectively. In brief, we calculate a blackbody continuum spectrum at $T_\mathrm{eff} = 10000$ K between 1400 and 10000 $\text{\AA}$.  We then calculate the fraction of the blackbody flux within the GALEX FUV and NUV bands, and then scale by an assumed ratio of continuum to bolometric flux of 0.6 following \citet{2015ApJ...809...79O}.  See the Python notebook associated with the online version of this paper for details on the calculations.  However, we emphasize that these are approximate corrections, and for FUV the correction factor is even more uncertain: the calculation assumes the dominant flux contribution is from blackbody radiation within the bandpass, assumes a flare blackbody temperature of 10000 K, and does not take into account the transmission across the bandpass.  Significant emission lines from C IV, He II, and Al II mean that the emission in the GALEX FUV bandpass likely includes non-blackbody contributions.  Improvements to the bolometric conversion would require spectra obtained during the flares to model the emission lines in the FUV and NUV bandpasses.  The assumption of a ``typical'' blackbody temperature of 10000 K for a flare continuum may not be appropriate in all cases given, e.g., the recent HST observation of a flare with an estimated color temperature of 40000 K \citep{2019ApJ...871L..26F}.

\begin{table}[ht]
\begin{threeparttable}
\caption{Summary of Detected Flares From The GJ 65 System}
\begin{tabular}{rrcrccr}
\toprule
 Flare &  Visit &  NUV Peak Time &  Duration &     log($E_{NUV}$) &      log($E_{FUV}$) &  NUV Strength\tnote{a} \\
  &   &  (UTC) &  (min.) &   (erg)   &  (erg)     &   \\
     1 &      1 &  2005-11-18 06:54:45 &       9.5 &  $29.45 \pm 27.92$ &  $28.86 \pm 27.84$ &      8.498467 \\
     2 &      2 &    peak not measured &       5.5 &  $28.95 \pm 27.80$ &  $28.00 \pm 27.64$ &      1.761480 \\
     3 &      2 &  2005-11-18 08:35:49 &      10.5 &  $28.88 \pm 27.89$ &  $28.34 \pm 27.81$ &      1.293150 \\
     4 &      3 &  2005-11-18 11:54:02 &      12.0 &  $29.17 \pm 27.97$ &  $28.64 \pm 27.89$ &      1.397152 \\
     5 &      3 &  2005-11-18 11:48:02 &       7.5 &  $28.95 \pm 27.86$ &  $28.52 \pm 27.80$ &      1.660546 \\
     6 &      4 &  2005-11-18 18:22:56 &       5.0 &  $29.42 \pm 27.88$ &  $29.02 \pm 27.83$ &      3.686975 \\
     7 &      5 &  2005-11-18 20:01:01 &       5.0 &  $28.73 \pm 27.77$ &  $27.91 \pm 27.67$ &      1.329063 \\
     8\tnote{b} &      5 &  2005-11-18 20:12:01 &      18.0 &  $31.31 \pm 28.73$ &  $30.80 \pm 28.66$ &    150.754465 \\
     9 &      6 &  2005-11-19 00:56:50 &       5.5 &  $28.73 \pm 27.83$ &  $28.53 \pm 27.81$ &      1.194350 \\
    10\tnote{c} &      7 &  2005-11-19 05:52:08 &       6.0 &  $29.48 \pm 27.97$ &  $28.23 \pm 27.77$ &      5.832215 \\
    11 &      7 &  2005-11-19 05:54:38 &       7.5 &  $29.05 \pm 27.94$ &  $28.52 \pm 27.88$ &      1.253377 \\
    12 &      8 &    peak not measured &       6.0 &  $28.98 \pm 27.91$ &  $27.80 \pm 27.75$ &      1.192360 \\
    13 &      8 &  2005-11-19 09:17:50 &      10.5 &  $29.39 \pm 28.05$ &  $28.81 \pm 27.97$ &      1.662115 \\
    14 &      9 &  2005-11-19 23:57:14 &       6.5 &  $29.06 \pm 27.81$ &  $28.50 \pm 27.73$ &      2.524152 \\
    15 &      9 &  2005-11-19 23:58:14 &       7.0 &  $29.31 \pm 27.88$ &  $28.88 \pm 27.83$ &      3.180030 \\
\end{tabular}
\begin{tablenotes}\footnotesize
\item [a] This is computed as $\mathrm{\left(p - 3\sigma_{p}\right) / i}$, where $\mathrm{p}$ is the count rate at the flare's peak and $\mathrm{i}$ is the count rate at the INFF level, and is a measure of the strength of the flare detection.
\item [b] This flare exceed the local non-linearity threshold of the GALEX detectors during the flare. The result is that the true flux densities are underestimated here.
\item [c] This flare is heavily truncated in both bands at the start of the visit, nearly entirely in FUV, thus the flare parameters are all lower limits at best. We do not include this event in the rest of our analysis, but list it here for completeness.
\end{tablenotes}
\end{threeparttable}
\label{flaretable}
\end{table}

\subsection{Alternative Flare Energy Estimation}\label{altflareest}
Several of the flare events identified by our algorithm do not exhibit a ``classical'' flare lightcurve morphology, wherein the flux starts near the INFF value and then undergoes a FRED-shaped evolution before returning to the INFF.  Multi-peaked flare events could be a single flare with a complex morphology (``complex''), or they could be composed of several flares happening concurrently on the stars' surfaces (``compound''). Some previous work has suggested that complex flare shapes are common at low energies (\cite{brasseur2019}, \cite{howard2021}). Low energy flares are also expected to be more common, which would result in more compound flares. Our detection approach described above and summarized in Table \ref{flaretable} defines a single flare event as one bound by a return to the INFF flux level, regardless of the shape. We were not able to come up with an alternative approach to flare detection---given only the data in this sample---that could distinguish whether any given flare event was classical, compound, or complex.

\begin{figure}
\includegraphics[width=0.8\textwidth]{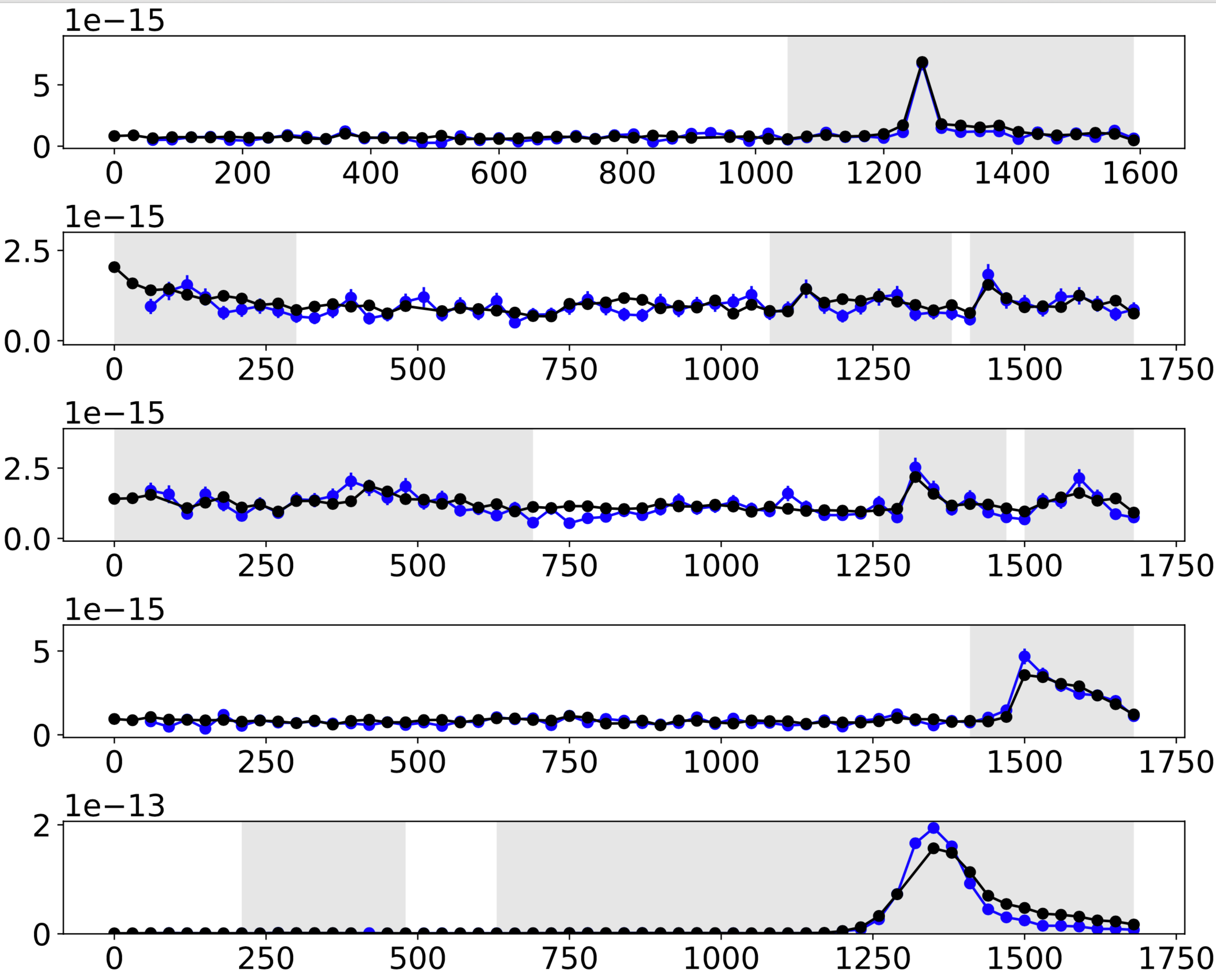}
\caption{Visit-level light curves for Visits 1-5 (top to bottom) with flare ranges split manually (Visits 2 and 3).  FUV is in blue, NUV is in black, and detected flare events are denoted by the grey backgrounds. \label{allflares_1_byhand}}
\end{figure}

\begin{figure}
\includegraphics[width=0.8\textwidth]{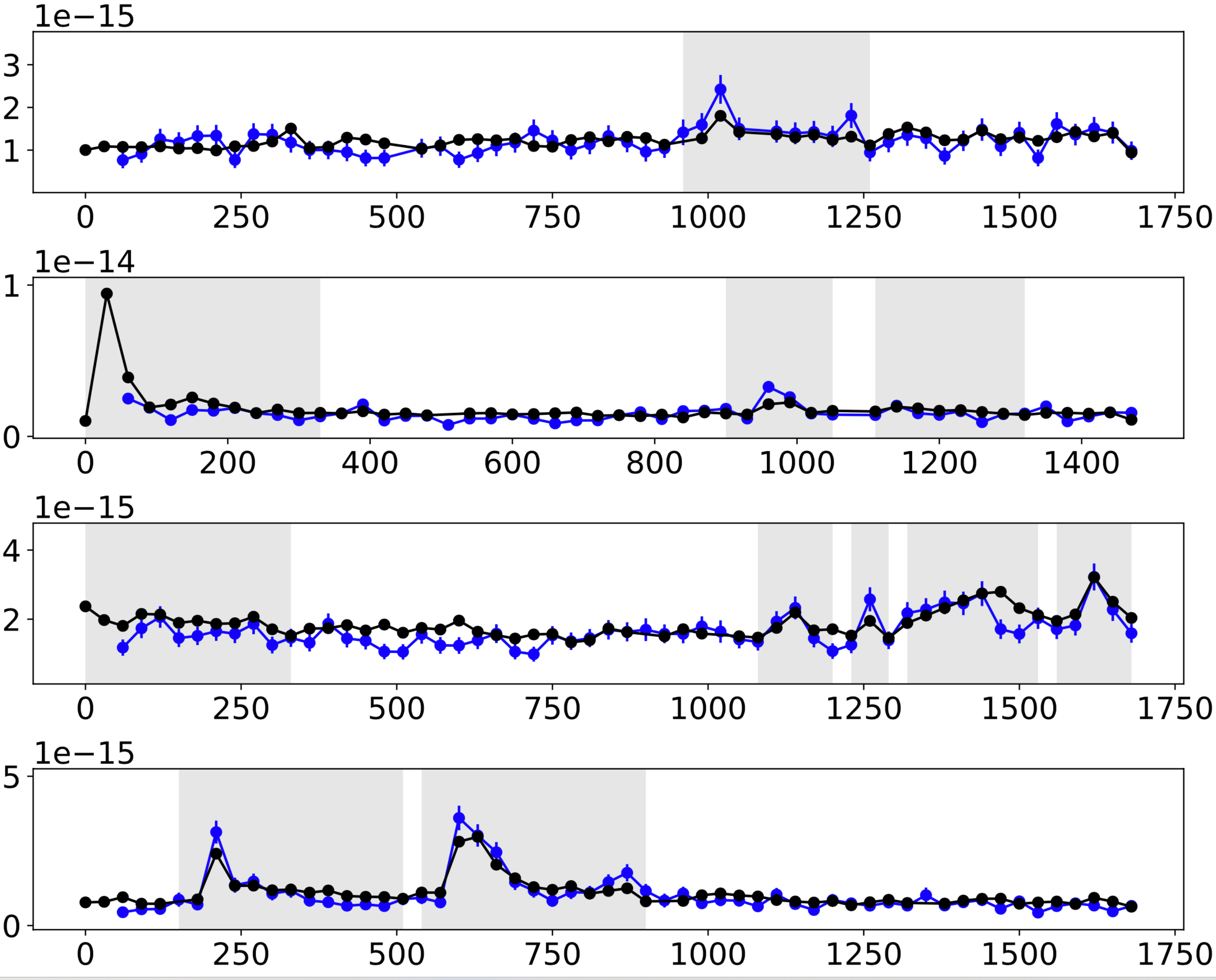}
\caption{Visit-level light curves for Visits 6-9 (top to bottom) with flare ranges split manually (Visits 7 and 8).  FUV is in blue, NUV is in black, and detected flare events are denoted by the grey backgrounds.  Note that Visit 9 was already manually split in the code as part of automated algorithm-based flare detection.  We did not split the second ``bump'' in the event in Visit 5 into a separate flare because the NUV flux does not show significant variability above the INFF value, which it should for real flare events since count rates are higher in the NUV. We also did not split the second ``bump'' in the second event in Visit 9 because the NUV flux is within 1-$\sigma$ of the neighboring fluxes and thus is not an obvious new peak. \label{allflares_2_byhand}}
\end{figure}

To explore the possibility that multi-peaked flares are caused by multiple, individual flares happening concurrently, we have examined each flare event detected by our algorithm for multiple peaks and manually split them into smaller ``flares'' (Figs. \ref{allflares_1_byhand} and \ref{allflares_2_byhand}). A summary of the energies of flares extracted in this manner are presented in Table \ref{flaretable_byhand} and are used to generate a second estimate of the flare frequency at low energy in Section \ref{sec:cite}.  Some groups have used Bayesian Markov Chain Monte Carlo (MCMC) methods to split multi-peaked flare events into individual flares \citep{2020AJ....159...60G, 2021ApJ...912...81K}.  These methods are promising, but still need to make assumptions about the underlying shapes of the individual flares when fitting the multi-peaked events.  To match how energies are calculated when we treat them as single flares, we estimate flare energies using the area under the curve for each piece of the event.  Note that flare numbers referenced in this document and in figures refer to those in Table \ref{flaretable} throughout.  To avoid confusion, we append these manually subdivided flares with a 'b' in Table \ref{flaretable_byhand}.

\begin{table}[ht]
\begin{threeparttable}
\caption{Summary of Flares When Split By Hand}
\begin{tabular}{rrcrcc}
\toprule
Flare &  Visit &  NUV Peak Time &  Duration &     log($E_{NUV}$) &      log($E_{FUV}$) \\
  &   &  (UTC) &  (min.) &   (erg)   &  (erg) \\
     1b &      1 & 2005-11-18 06:54:45 &       9.5 & $29.45 \pm 27.92$ & $28.86 \pm 27.84$ \\
     2b &      2 &   peak not measured &       5.5 & $28.95 \pm 27.80$ & $28.00 \pm 27.64$ \\
     3b &      2 & 2005-11-18 08:30:49 &       5.5 & $28.59 \pm 27.75$ & $27.92 \pm 27.67$ \\
     4b &      2 & 2005-11-18 08:30:19 &       5.0 & $28.56 \pm 27.72$ & $28.12 \pm 27.67$ \\
     5b &      3 & 2005-11-18 11:54:02 &      12.0 & $29.17 \pm 27.97$ & $28.64 \pm 27.89$ \\
     6b &      3 & 2005-11-18 11:48:02 &       4.0 & $28.69 \pm 27.72$ & $28.28 \pm 27.67$ \\
     7b &      3 & 2005-11-18 11:48:32 &       3.5 & $28.61 \pm 27.69$ & $28.14 \pm 27.62$ \\
     8b &      4 & 2005-11-18 18:22:56 &       5.0 & $29.42 \pm 27.88$ & $29.02 \pm 27.83$ \\
     9b &      5 & 2005-11-18 20:01:01 &       5.0 & $28.73 \pm 27.77$ & $27.91 \pm 27.67$ \\
    10b\tnote{a} &      5 & 2005-11-18 20:12:01 &      18.0 & $31.31 \pm 28.73$ & $30.80 \pm 28.66$ \\
    11b &      6 & 2005-11-19 00:56:50 &       5.5 & $28.73 \pm 27.83$ & $28.53 \pm 27.81$ \\
    12b\tnote{b} &      7 & 2005-11-19 05:52:08 &       6.0 & $29.48 \pm 27.97$ & $28.23 \pm 27.77$ \\
    13b &      7 & 2005-11-19 05:53:08 &       3.0 & $28.66 \pm 27.73$ & $28.38 \pm 27.69$ \\
    14b &      7 & 2005-11-19 05:52:08 &       4.0 & $28.67 \pm 27.78$ & $27.80 \pm 27.70$ \\
    15b &      8 &   peak not measured &       6.0 & $28.98 \pm 27.91$ & $27.80 \pm 27.75$ \\
    16b &      8 & 2005-11-19 09:09:50 &       2.5 & $28.36 \pm 27.69$ & $27.88 \pm 27.61$ \\
    17b &      8 & 2005-11-19 09:09:20 &       1.5 & $27.91 \pm 27.56$ & $27.83 \pm 27.52$ \\
    18b &      8 & 2005-11-19 09:11:20 &       4.0 & $29.11 \pm 27.86$ & $28.52 \pm 27.78$ \\
    19b &      8 & 2005-11-19 09:09:50 &       2.5 & $28.92 \pm 27.75$ & $28.27 \pm 27.67$ \\
    20b &      9 & 2005-11-19 23:57:14 &       6.5 & $29.06 \pm 27.81$ & $28.50 \pm 27.73$ \\
    21b &      9 & 2005-11-19 23:57:44 &       6.5 & $29.31 \pm 27.86$ & $28.88 \pm 27.81$ \\
\end{tabular}
\begin{tablenotes}\footnotesize
\item [a] This flare exceed the local non-linearity threshold of the GALEX detectors during the flare. The result is that the true flux densities are underestimated here.
\item [b] This flare is heavily truncated in both bands at the start of the visit, nearly entirely in FUV, thus the flare parameters are all lower limits at best. We do not include this event in the rest of our analysis, but list it here for completeness.
\end{tablenotes}
\end{threeparttable}
\label{flaretable_byhand}
\end{table}

\section{Flare Analysis}\label{sec:flaredesc}
Detailed visit-level light curves with zoomed sections centered on each detected flare are presented in Appendix \ref{app_flareplots}, Figs. \ref{flarefig1}, \ref{flarefig2}, \ref{flarefig3}, \ref{flarefig4}, \ref{flarefig5}, \ref{flarefig6}, \ref{flarefig7}, \ref{flarefig8}, and \ref{flarefig9}.  Each light curve is binned at 30 seconds, and we do not include any bins that contain less than 20 seconds effective exposure; when present, these short-exposure bins typically occurred at the beginning or ends of visits.  UV Ceti is known to be a very active system, and we detect at least one flare in every visit we analyzed in this work except the last (not shown in Appendix \ref{app_flareplots}). The flare events are all within the UV energy range $\sim28.5 < \log{E} < 29.5$ ergs with the exception of the largest flare (Flare 8 in Visit 5) previously detected by \citet{2017AJ....154...67M} at the visit-level. In the following sections, we estimate the flare frequency in the $28.5 < \log{E} < 29.5$ ergs range, where all but one of our detected flares reside, and check if this rate is consistent with extrapolations of FFDs at higher energies.  Flare 8 by itself is remarkable because of how bright it is, enabling a search for in-flare variability.

Flare 10 from Visit 7 is heavily truncated, where the peak likely occurred before the GALEX visit started.  The FUV channel started to collect data a little later than the NUV, which makes for so little coverage that it is essentially undetected in the FUV band. To make matters worse, the first NUV bin and first two FUV bins are impacted by low effective exposure times and are excluded as mentioned above. The first and last bins can sometimes be unreliable due to how gPhoton defines bin boundaries, which may not always align with the first or last valid photons for that particular visit.  Checking the time coverage for the first and last bins of gPhoton light curves is always recommended, and those with very little effective exposure times are best ignored. All these issues compound to make the identification of this variability as a flare questionable at best. Even if this were an actual flare event, the lack of coverage during the event would make the estimated energy and duration very poor lower limits. Because of these issues, we have decided to include this event as a possible flare detection in Table \ref{flaretable}, but we do not include it in our flare rate estimation.

\subsection{Flare Rate} \label{sec:cite}
\subsubsection{Flare Selection}
Without spatially resolving a star's surface, it can be difficult to determine if a given flux increase is composed of one flare event, multiple distinct flare events, or a chain reaction of physically related flare events. It can also be difficult to say with certainty that a particular event is a stellar flare when the event has an unusual shape, or the ratio of peak flux to INFF is relatively small. For this paper, we prefer to define a ``flare'' as variations in stellar flux density that trigger our flare search criteria.  In this case, a candidate flare is only included in the flare rate calculation if both the INFF and local maximum, assumed to be the flare peak, were observed, as well as some baseline before and after the peak to capture the rise to, and decay from, peak flux. These criteria were chosen because the INFF, peak flux, and slopes of the rise and decay phases are the minimum parameters required to model a classical flare light curve.  Some of the flares that pass these tests do not return to the INFF value before the end of the visit, or have pre-flare components that are above the INFF.  The derived energies in these cases are lower limits, but since in these cases the peak flux and a majority of the event are captured, the majority of the flare energy should also be captured in the GALEX data.  However, compared to other flare surveys with much longer continuous observing windows, the limited time baseline of each GALEX visit makes it more difficult to estimate when a flare starts and ends.  These criteria do not apply to the alternative method, where a flare event is manually split into individual flares (per Section \ref{altflareest}); by construction, in that case, the flux from individual flares has not returned to the INFF value.

When selecting flares to use for the flare frequency estimation from Table \ref{flaretable}, the following flare events are excluded.  As previously described, Flare 10 is a questionable flare event to begin with, and is heavily truncated even if it is a flare, and thus is excluded from the flare rate analysis.  In addition, Flare 2 and Flare 12, where the local maximum occurs at the beginning of the visit, are excluded.  In these cases, the GALEX visit only captured the decay phase of a (presumably) larger flare. Light curves with local maxima at the end of the visit would also have been excluded, although none are present in these data. We have chosen to decompose the light curve in Visit 9 into two distinct, ``fast rise, exponential decay'' (FRED)-shaped flares (Flares 14 \& 15). Although Flare 3 looks like it may also be two distinct FRED-like shapes, only one of the peaks passes our search requirement of being $3\sigma$ above the INFF simultaneously in both bands, so we decided not to decompose this event. Many of the other flares are complex events, with several local maxima, but we have chosen to not attempt to partition these into individual flares because they are not clearly composed of a finite number of ``classical FRED'' shapes. Any flare rate calculation will depend heavily on the criteria defined to identify distinct flare events.  This leaves a total of 11 flares to use for estimating the flare frequency in the energy bin $28.5 \leq \log{E} \leq 29.5$.

When selecting flares to use for the alternative flare frequency estimation from Table \ref{flaretable_byhand}, the following flare events are excluded.  Flares 2b and 15b correspond to Flare 2 and 12, and are excluded for the same reason.  Flare 10b corresponds to Flare 8 and is excluded because it is the one large flare outside our energy bin we are considering.  Flare 12b corresponds to Flare 10 that is heavily truncated, and is excluded for the same reason.  Finally, Flares 16b and 17b have estimated flare energies below the bin we are considering, and thus are excluded.  This leaves a total of 15 flares to use for estimating the flare frequency when breaking down flare events manually.

\subsubsection{Flare Recovery Testing}
To determine the fraction of flares that may have been missed by our flare search algorithm, we conducted a recovery test.  We create 100,000 simulated light curves containing flares using the \citet{2014ApJ...797..122D} flare template model.  For each simulated light curve, we inject one, two, or three flares within the visit.  A baseline count rate for a source at an NUV magnitude of 18 is assumed, which is the approximate brightness of GJ 65 in the GALEX NUV band. To match the GJ 65 light curves here, our simulated light curves use time bins of 30 seconds.  We use a detection threshold of three sigma to match our detection criteria here, but do not generate both FUV and NUV light curves, since all of the detected flares for GJ 65 have FUV to NUV flux ratios very close to unity.  We then randomly select from uniform distributions, for each flare: a peak flare magnitude ranging from NUV of 13-18, a start time during the central 1400-second observation baseline (avoiding the first and last 200 seconds of the observing window), and a flare full width at half maximum ranging from 1 to 300 seconds. Each simulated light curve is then run through our flare detection algorithm.

Detected flare events are compared with the injected flares.  We define a few categories for each flare event.  An injected flare is considered ``detected'' if the peak of an event detected by our algorithm falls within two time bins of the injected flare's peak, and the calculated energy of the detected event is within a factor of five of the injected flare's energy.  We choose this tolerance factor because we are estimating the flare frequency in an energy bin that covers a full order of magnitude ($28.5 \leq \log{E} \leq 29.5$).  If a detected event's peak is within two time bins of an injected flare's peak, but the energy is not within a factor of five, then it is counted as ``detected with wrong energy''.  This is commonly the case when a large flare is near one or more smaller flares: the time of peak flux matches the time of peak flux in an injected flare, but our algorithm counts them all as one event and thus derives a total energy that is far off from the injected energy.  Any injected flares that do not have times of peak flux near any detected flare events are counted as ``missed''.  This category can include those smaller flares that occur near a larger flare within the visit.  A final category of ``false positive'' is reserved for those flares that our detection algorithm identifies but do not match the times of any injected flares.

\begin{figure}
\includegraphics[width=0.95\textwidth]{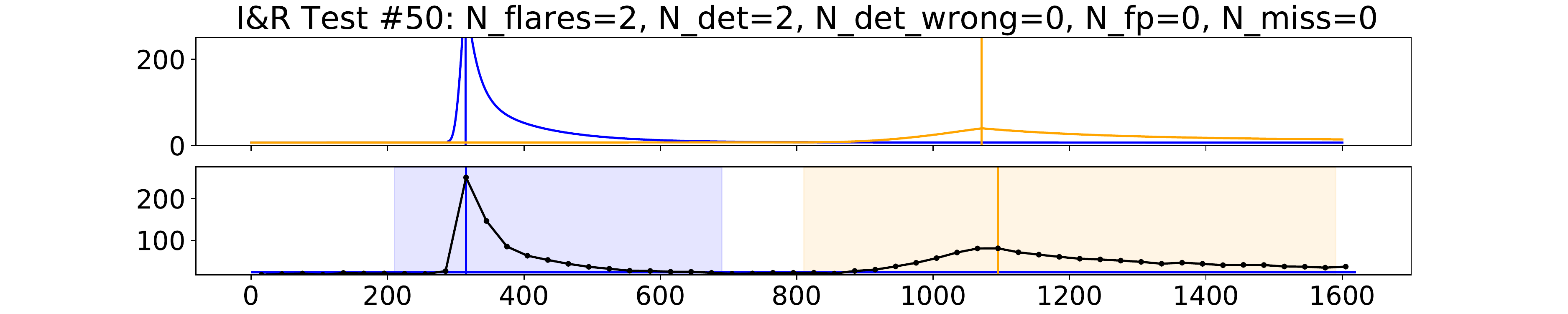}
\caption{Example of a simulated visit light curve containing two flares (peaks marked with vertical lines in the top panel.) In this case, the flares are spaced within the visit such that our algorithm correctly identifies them as individual flares (blue and orange shaded regions in the bottom panel), and the energies match the injected flare energies within our tolerance factor (a factor of 5.) \label{flare_050}}
\end{figure}

\begin{figure}
\includegraphics[width=0.95\textwidth]{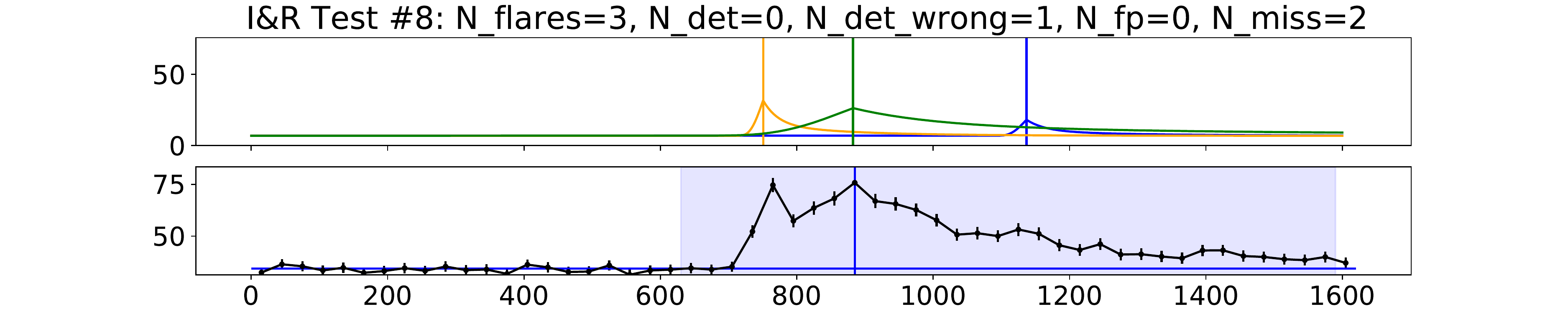}
\caption{Example of a simulated visit light curve containing three flares (peaks marked with vertical lines in the top panel.) Our algorithm identifies this as a single ``flare event'' (blue shaded region in bottom panel.) The peak time of the detected flare event matches the peak time of the strongest injected flare, but the derived energy is larger than our tolerance factor (a factor of 5) since multiple injected flares are contributing flux. \label{flare_008}}
\end{figure}

Figs.\ \ref{flare_050} and \ref{flare_008} show sample visit light curves with injected flares.  Fig.\ \ref{flare_050} provides a sample of a light curve with two injected flares well-separated in time.  In this case, the flux returns to the INFF value before the next flare occurs, thus, our algorithm is able to correctly identify two individual flares in this visit, and the estimated energies are within our tolerance factor to count as being ``detected''.  Fig.\ \ref{flare_008} provides a sample of a light curve with three injected flares that occur close enough in time such that the flux level never returns back to the INFF level before the other flares occur.  This means that our flare detection algorithm treats this as a single ``flare event'', rather than the three ``classically'' shaped flares that were injected.  While many flares exhibit the FRED-shaped model, not all flares do, especially at low energies in the ultraviolet \citep{brasseur2019}. In these cases, is a given flare event composed of a series of FRED-shaped flares occurring concurrently, or a single flare event that does not follow the FRED-shape morphology?  Answering that question is beyond the scope of this work, but it does impact our estimation of a flare frequency.  Thus, we estimate flare energies in two ways: following our algorithm that defines a single flare event based on a return of the flux to the INFF value, and one where we force energy calculations by splitting any multi-peaked flare events into individual flares.  These two methods cover the two cases: in the former, the assumption made is that all flare events are from a single flare with a complex shape, while in the latter, the assumption made is that every flare event with more than one peak is composed of individual flares happening concurrently.

\begin{figure}
\includegraphics[width=0.8\textwidth]{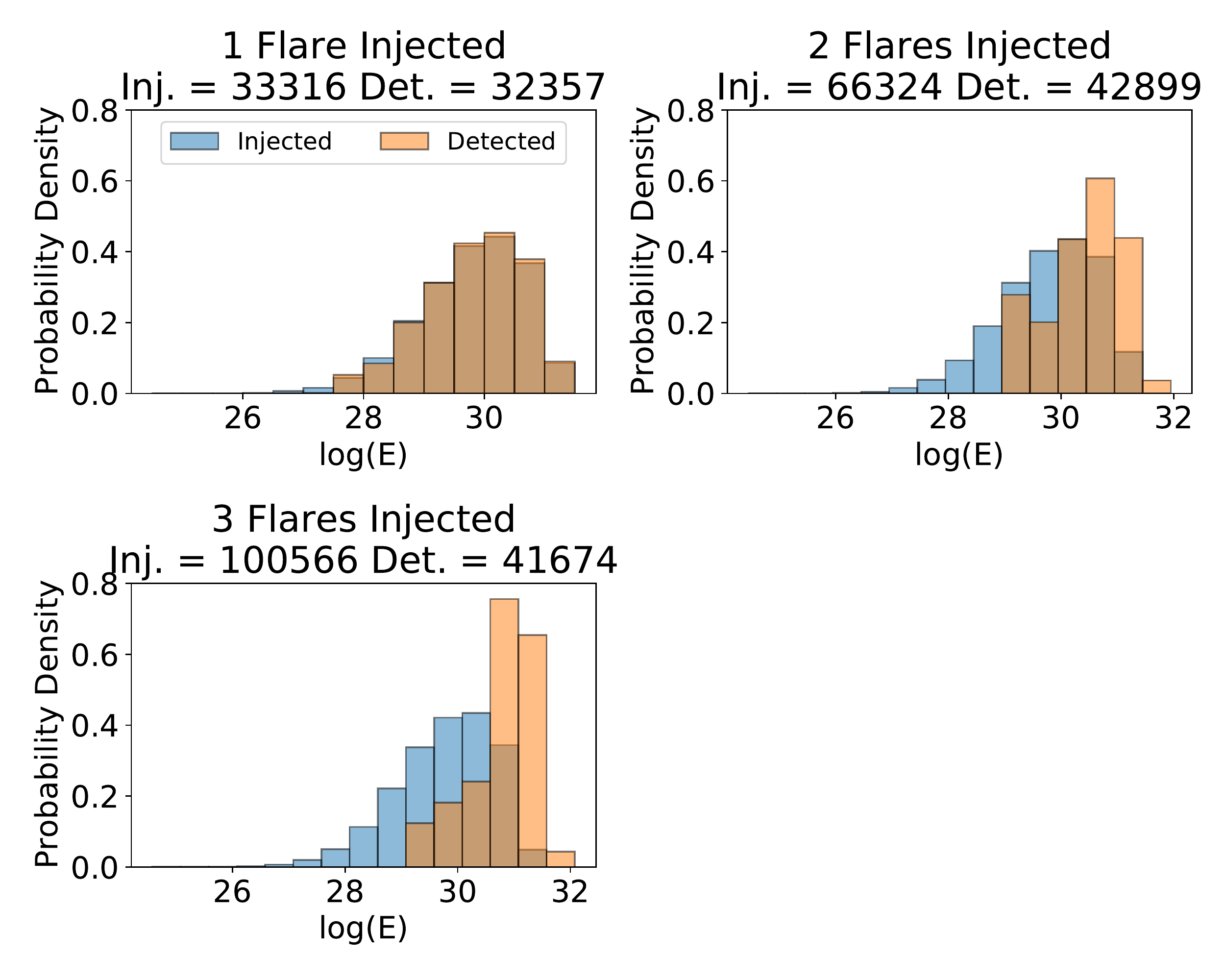}
\caption{Comparison of the distributions of flare energies injected into the light curves (blue) and those detected and estimated by our flare algorithm (orange).  Top-left: flares in light curves with only one flare injected.  Top-right: flares in light curves with two flares injected into the visit.  Bottom-left: flares in light curves with three flares injected into the visit. The number of flares injected and detected are reported in the plot titles for all three cases. \label{flare_inj_det}}
\end{figure}

To assess the impact of derived flare energies when there are multiple flares occurring concurrently, we compare the distribution of injected flare energies with the distribution of detected flare energies for the cases of one, two, and three injected flares per visit.  False positive flare detections were less than 10\% the total number of injected flares in all three cases.  Fig. \ref{flare_inj_det} compares these distributions, using a bin size of log(E) = 0.5, for the injected flares (blue) and detected flares (orange).  In visits where there is a single injected flare, the estimated energies match well with the injected energies, and the vast majority ($\sim 94\%$) of the flares are detected with energies that match the injected flare's energy within a factor of five.  For visits that contained two injected flares, only $\sim 43\%$ of the injected flares were detected with energies within a factor of five, while $\sim 45\%$ of the flares were missed entirely.  For visits containing three injected flares, only $\sim 25\%$ of the flares were detected with energies within a factor of five, and $\sim 63\%$ of the flares were missed entirely.

To further complicate the matter, GJ 65 is a binary system of two M dwarfs, and the individual sources are not resolved by GALEX, so we are unable to assign any specific flare to UV Ceti or BL Ceti.  The two components have similar spectral types and rotation rates, but very different magnetic field behaviors and morphologies. Although UV Ceti is the more active component, we are limited to estimating a flare rate for the GJ 65 system as a whole.   Rather than attempt to assign some fraction of the events to each star, we bound our flare frequency estimation by assuming 100\% of the events come from one of the stars, and then by assuming the two stars split the events equally.  Thus we calculate an upper and lower estimate for the flare frequency.

We estimate the flare frequency between $28.5 \leq \log{E} \leq 29.5$ by dividing the total number of non-truncated flares detected within this energy range (11 using our algorithm, 15 when split by-hand), by the total observation time on source (15682 seconds).  This yields an estimated flare frequency of $\log{\#/\mathrm{hour}} = 0.40 \pm 0.11$ for the upper-bound case (where all flares come from one of the stars in the system), and a lower-bound flare frequency (where the detected flares are split evenly) of $\log{\#/\mathrm{hour}} = 0.10 \pm 0.15$.  When split by hand into smaller flares, the estimated flare frequencies are $\log{\#/\mathrm{hour}} = 0.54 \pm 0.10$ for the upper bound case and $\log{\#/\mathrm{hour}} = 0.24 \pm 0.14$ for the lower-bound case.  The quoted estimated uncertainties are $1\sigma$, derived from standard counting statistics.  Given the relatively large additional uncertainties introduced by the assumptions made, we do not apply the approximations for small-number counting statistics to further refine the confidence intervals \citep{1986ApJ...303..336G,2003MNRAS.340.1269E}.  The flare frequency estimation is available as part of the Python notebooks associated with the online version of this paper.

\begin{figure}
\includegraphics[width=0.8\textwidth]{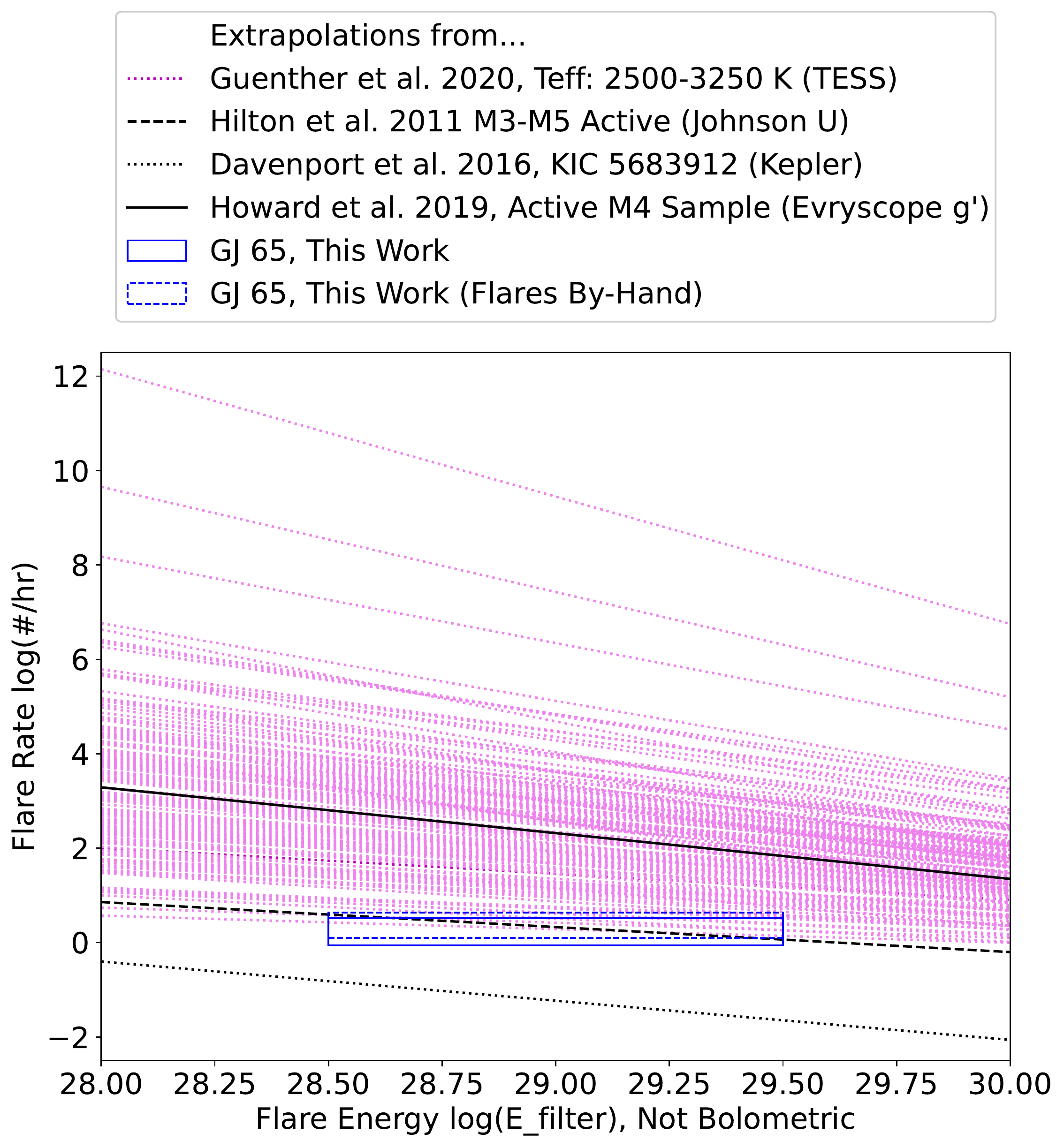}
\caption{Blue rectangular region: Estimated flare frequency based on detected flares from the GJ 65 system assuming they all come from one star (upper bound) or are split equally between the two (lower bound). The width of the region is bounded by the approximate lowest and highest energies detected in our flare sample (excluding Flare 8). The dashed blue box represents the flare frequency estimate when breaking flare events into smaller flares by hand.  Other FFD from the literature are shown for comparison. No conversion between the energies to account for the different filters is applied, since between the NUV and optical none would change the derived bolometric energies by more than an order of magnitude compared to each other.\label{ffdfig}}
\end{figure}

As an exercise, we can compare this estimated rate with the values expected when extrapolating from other recent FFD calculations for M dwarfs.  By no means is our estimated flare rate a strong constraint on flare rates for mid-M dwarfs at these low energies as a population, but it is still instructive to see if our estimated flare frequency is in-line with results from much larger surveys.  There are a variety of sources we compare against.  We use the ``M3-M5 Active'' FFD from \citet{2011PhDT.......144H}, the lone M star smaller than $0.2 \; \mathrm{M_\odot}$ from \citet{2016ApJ...829...23D}, V541 Lyr = KIC 5683912 (although it may not be a dwarf based on spectroscopic data from \citet{2016A&A...594A..39F}), the ``M4 Active'' sample from \citet{2019ApJ...881....9H}, and the M dwarf FFDs from \citet{2020AJ....159...60G} for TESS targets with $2500 \leq \mathrm{T_{eff}} \leq 3250$ K.  Fig.\ \ref{ffdfig} shows our estimated flare frequencies for GJ 65 compared to FFD from these various sources based on detected flares of higher energies and different optical passbands.  Note that we do not attempt to correct them all to bolometric luminosities to account for differences in the filters, since in the NUV and optical none would change the bolometric luminosity by more than an order of magnitude compared to any other, and thus would have minimal impact on the log-log relation being compared.  Despite the approximations and uncertainties in the analysis, the estimated flare frequency based on these low-energy GALEX flares from the GJ 65 system is comparable to the predicted values based on extrapolating these other FFD determinations for similar M dwarfs down to these energies.  Breaking down flare events into smaller flares (dotted blue box in Fig.\ \ref{ffdfig}) does not change the estimated flare frequency significantly, since the number of flares in this energy bin only increases from 11 to 15.

\subsection{Detailed Analysis of Flare 8}\label{flare8}
The eighth flare in our sample is the largest by orders of magnitude in both peak flux and integrated energy, so it allows for much more detailed analysis.  We generate NUV and FUV light curves with 5-second bins for this visit, to better study the evolution and shape of the flare.  The GALEX microchannel plate detectors have two sources of instrumental non-linearity: a global non-linearity caused by the time it takes the electronics to assemble photon lists, and a local non-linearity caused by limited current supply to small regions near bright sources.  For a detailed discussion, refer to \citet{2007ApJS..173..682M}, Section 4.4.  The global non-linearity can be corrected to better than 10\% through calibration, but the local non-linearity is not reliably corrected.  \citet{2007ApJS..173..682M} conducted an analysis of bright white dwarf standards, and identified count rates where the drop-off exceeds 10\% (gPhoton uses a threshold of 109 cps in FUV and 311 cps in NUV).

Flare 8 is bright enough to induce non-linear detector response.  The NUV rate exceeds the 10\% local non-linearity threshold for $\sim$75 seconds on either side of the flare's maximum brightness (maximum count rate in NUV is $> 800 $ cps).  From the predicted-to-measured count rates of bright white dwarfs presented in \citet{2007ApJS..173..682M}, the NUV count rate during peak flare brightness would be suppressed by at least 30\% (albeit with a large uncertainty).  In the FUV the 10\% threshold is only exceeded for $\sim$25 seconds on either side of peak flare brightness, but at the maximum FUV count rate ($\sim$159 cps) the count rate suppression would still be at the 25\% level based on the approximate formula given in \citet{2007ApJS..173..682M}.  These estimates, based on Figure 8 from \citet{2007ApJS..173..682M}, are highly uncertain: the white dwarf calibrators show significant scatter around the fitted relation, even between measurements at a single count rate (at an expected count rate of 800 cps, which coincidentally matches the peak NUV count rate during the GJ 65 flare, the measured count rates range from $\sim$300-800 cps).  The local non-linearity is known to depend on the change in gain across the field-of-view, which is not well mapped using the relatively small sample of targeted, bright white dwarf calibrators that typically fell in the centers of the fields-of-view, and the timestamps of the observations are not provided, and there have been no investigations on whether the impact of local non-linearity changed over the course of the mission.

Without recognizing that the count rates are in the non-linear regime, one would derive a significant FUV-NUV color difference, and in fact, an FUV to NUV ratio that changes during the flare itself.  The time-evolution of the FUV to NUV ratio during a flare can place strong constraints on the flare mechanism and properties, especially on the effective temperature of the flare's blackbody component \citep{2003ApJ...597..535H}.  Even in the absence of time-resolved spectra to study how individual lines change during a flare, one could use information about the emissivity and the characteristics of the flare to deduce some information.  For flares with QPP detected in two bands, one can compare the strengths of the detected signals with the fluxes in both bands.  A QPP signal strength that does not scale with the fluxes (i.e., signal-to-noise) from those bands is evidence that the signal is approximately wavelength-independent across the bandpasses.

The flare spectral energy distribution consists of isolated emission lines and a hot blackbody continuum feature.  Different species have line emission contribution in the FUV bandpass compared to the NUV, with very different formation temperatures: in the FUV bandpass it is predominantly C IV and Si IV, which form at temperatures near 100000 K, while in the NUV bandpass it is Mg II, with formation temperature $\lesssim$ 30000 K \citep{2016ApJ...830..154F}.  A QPP signal that is stronger in the GALEX FUV than the GALEX NUV is evidence that it arises from emission lines that dominate the FUV bandpass, while a signal that is wavelength-independent or dominant in the NUV is evidence that the QPP originates in the hot blackbody emission, especially since the GALEX bandpass has little sensitivity near the Mg II lines and is dominated by continuum emission.  However, without an accurate correction (and associated uncertainties to that correction) for local non-linearity to account for the missing flux, such an analysis is not possible for Flare 8.  While the branch of gPhoton we used for this paper does not output quality flags as part of the light curves, the main branch of gPhoton does, and readers are reminded to always pay attention to these quality flags, which include checks on count rates that exceed the 10\% local non-linearity threshold.

\subsubsection{Quasi-periodic Pulsations}
While the effects of local non-linearity limit our ability to study the color evolution of the flare, it does not preclude a search for QPPs.  Detecting a consistent signal in both the FUV and NUV bands is a good indicator: unless the signal is strongly correlated with the dither pattern, there is nothing in the GALEX hardware design that would suggest a signal would appear in both bands by chance.  \citet{2018ApJS..238...25D} conducted an extensive test of gPhoton light curves to analyze false positives using several thousand bright, blue sources, and found that false positives are nearly always correlated with the dither pattern of the telescope.  Thus, any signal detected in both bands that is not correlated with position on the detector (in gPhoton, the ``detrad'' parameter) is a strong candidate for a real signal, even if there is missing flux due to local non-linearity.

The presence of in-flare variability is readily apparent even by visual inspection of Flare 8.  Such periodic variability could be caused by QPP \citep[and references therein]{1969ApJ...155L.117P,1974A&A....32..337R,2010SoPh..267..329K,2003A&A...403.1101M,2015SoPh..290.3625S,2016MNRAS.459.3659P}, which for stars other than the Sun, have been observed with periods ranging from a few seconds to hundreds of seconds.  In at least one case, multiple periods have been detected during a single flare \citep{2015ApJ...813L...5P}.  \citet{2006A&A...458..921W} used time-tagged GALEX photon events to detect QPP with periods of 30-40 seconds from large flares from AF Psc, GJ 3685A, Cr Dra, and SDSS J084425.9+513830.5, while \citet{2018MNRAS.475.2842D} used a wavelet analysis on gPhoton light curves to detect QPP in flares from six stars, including those in \citet{2006A&A...458..921W}.

As an initial quick check, we remove a smooth trend representing the unknown, aperiodic flare shape using a Savitzky-Golay filter.  The residuals are then run through a Lomb-Scargle periodogram, where a clear peak around 48 seconds is detected.  We also run a periodogram on a large time period of the light curve prior to the flare event, to test if there is intrinsic variability at this frequency, but find no statistically significant peaks.  To check if this is an instrumental effect, we generate 5-second cadence light curves for three of the closest known sources to GJ 65 (to check if there are any local systematics) and also three of the closest known sources within one magnitude of the peak brightness during the flare, in case the systematics only show up for very high count rates.  In no cases do we detect any statistically significant peaks in the periodograms. Note that a similar test can not be done using blank sky near GJ 65 itself, because the sky background in GALEX is typically very low. At the position of GJ 65, the estimated sky background is $\sim$0.59 cps within a 17.3\arcsec annulus.  To check for the presence of a $\sim$50 second signal, one would want to use at most 25 second bins for Nyquist sampling, resulting in $\sim$15 photon events per bin.  At these count rates, a signal at the 10\% level like that found on GJ 65 would be lost in the photon noise, thus tests must be done on known sources and not empty patches of sky.  Finally, we examine the frequency of the dither pattern to ensure this is not an induced signal caused by the motion of the spacecraft during the observation, and find no peak caused by the change of the spacecraft boresight that matches the period of the QPP.  All of these tests can be found in the Python notebooks associated with the online version of this paper.  Thus confident that this variability is unique to GJ 65 and only present during the flare itself, we conduct a more thorough analysis of the QPP.

\begin{figure}
\centering
\includegraphics[width=0.8\textwidth]{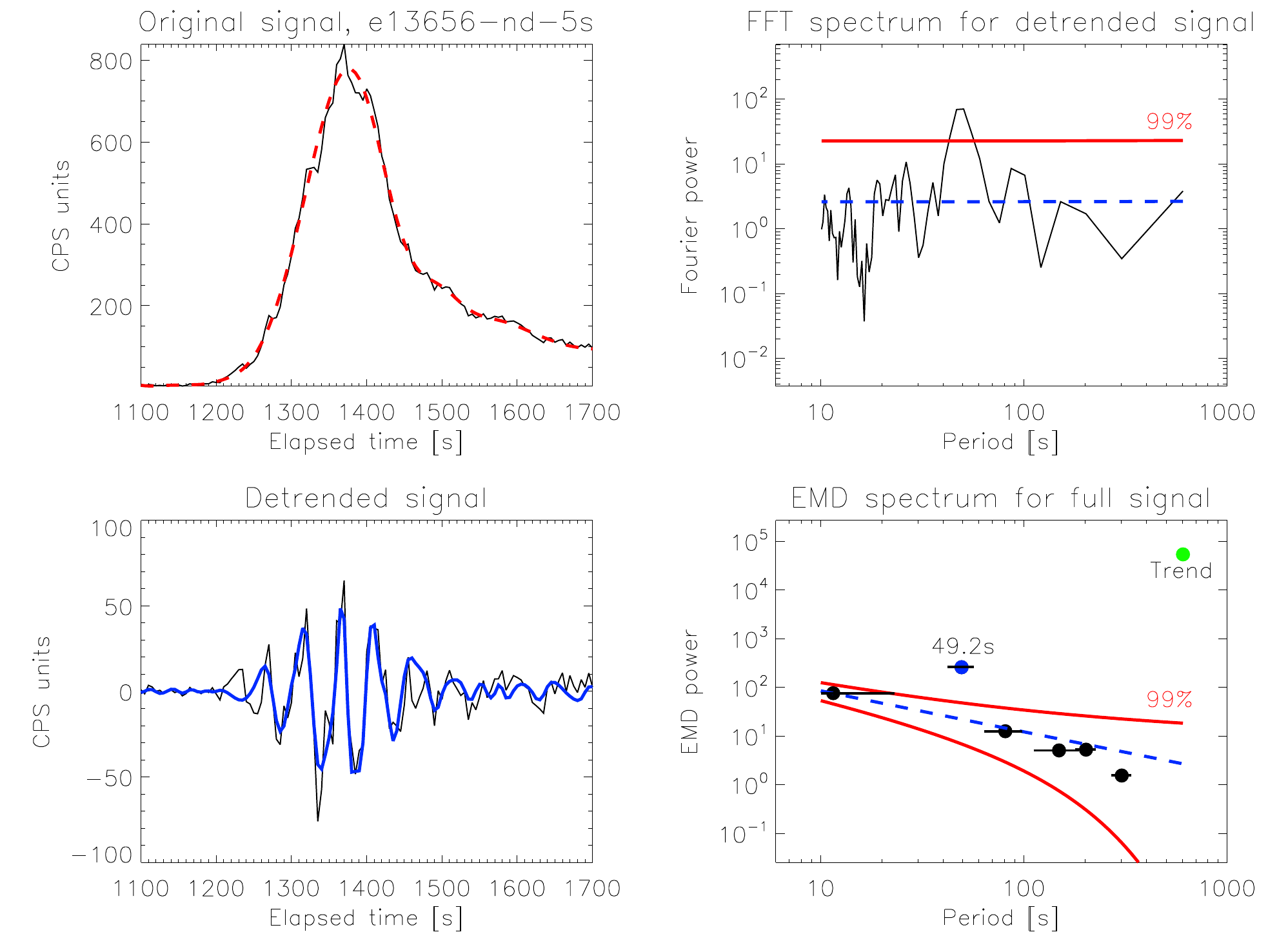}
\caption{QPP analysis of NUV light curve for Flare 8. Top left: Flare 8 light curve with the overall flare trend (the red dashed line), determined by the empirical mode decomposition (EMD).
Top right: Fourier power spectrum for the detrended flare light curve (see the bottom left panel), with the power-law fit (the blue dashed line) and the 99\% confidence level (the red solid line).
Bottom right: Dependence of the total power (circles) against mean period for each intrinsic oscillatory mode and flare trend determined in the original signal (see the top left panel) with EMD (the ``EMD power spectrum''). The blue dashed line shows the overall slope of the spectrum. The red solid lines show the corresponding 99\% confidence intervals.
Bottom left: The original flare light curve with the flare trend subtracted (the black line) and the oscillatory mode found to be statistically significant in the EMD analysis (the blue line). The elapsed time in all panels is relative to the start of the GALEX visit.
\label{emd_nuv}}
\end{figure}

\begin{figure}
\centering
\includegraphics[width=0.8\textwidth]{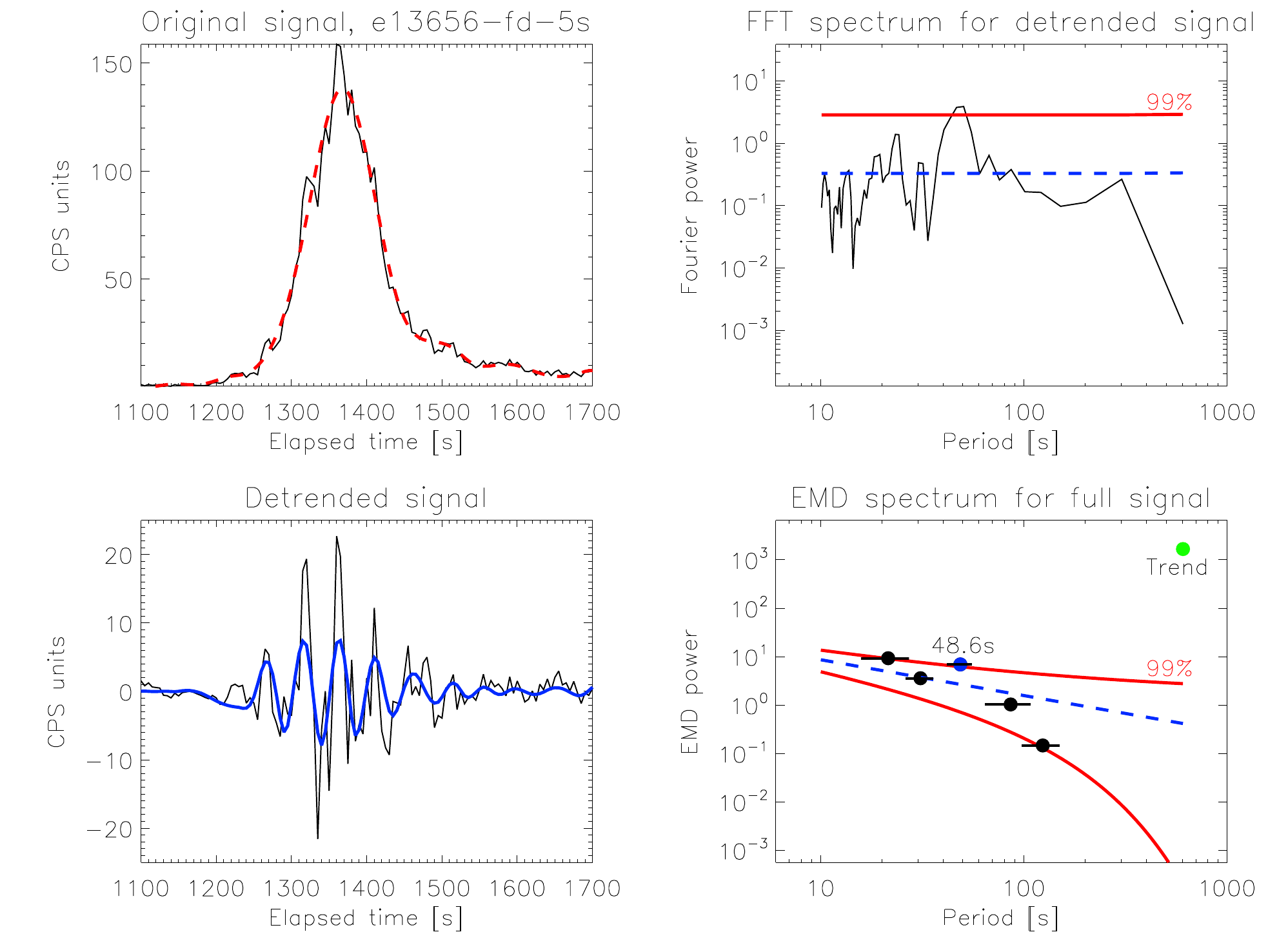}
\caption{QPP analysis of FUV light curve for Flare 8.  The panel layout is the same as in Fig.~\ref{emd_nuv}.
\label{emd_fuv}}
\end{figure}

We applied empirical mode decomposition \citep[EMD;][]{1998RSPSA.454..903H, 2008RvGeo..46.2006H} and Fourier transform methods\footnote{The IDL routines used for the analysis can be accessed via \url{github.com/Sergey-Anfinogentov/EMD_conf}. Note, the project is still under development so the interested reader is encouraged to contact the software authors (\href{mailto:d.kolotkov.1@warwick.ac.uk}{d.kolotkov.1@warwick.ac.uk}) directly for the most recent updates and guidance.} for detection and analysis of QPP in Flare 8 observed in the NUV and FUV bands. The use of two essentially independent methods for a raw signal decomposition into intrinsic oscillatory components and their analysis was motivated by the recent review of state-of-the-art techniques for detecting QPP in solar and stellar flares by \citet{2019ApJS..244...44B}, who suggested that confidence in the QPP detections can be improved if more than one detection method is employed. EMD allowed us to decompose the observed flare light curves into five (FUV) and six (NUV) intrinsic quasi-oscillatory modes and aperiodic trends (see Figs.~\ref{emd_nuv} and \ref{emd_fuv}). The flare aperiodic trends were determined by summing up all EMD-revealed timescales longer than half a length of the original signals. After removing the aperiodic trends, we calculated Fourier power spectra for both observational signals and estimated properties of the background noise and statistical significance of Fourier peaks in comparison with it, adapting methods from \citet{2005A&A...431..391V} and \citet{2017A&A...602A..47P}. The Fourier spectra for both signals are found to be fairly flat, indicating the absence of correlated noise in the observations.

We assessed the statistical significance of detected EMD modes following the scheme developed in \citet{2016A&A...592A.153K}. For this, we estimated the total energy (as a sum of all instantaneous amplitudes squared) and the mean period (using global wavelet analysis) of each individual EMD mode. In analogy with the Fourier power spectrum, the dependence between the EMD modal power and mean periods can be referred to as an ``EMD power spectrum''. The concept of an EMD power spectrum was previously used for the analysis of QPP in solar and stellar flares in e.g. \citet{2015A&A...574A..53K, 2018MNRAS.475.2842D, 2018ApJ...858L...3K, 2019MNRAS.482.5553J}. Using properties of noise detected in the Fourier analysis, we construct the upper and lower confidence intervals in the EMD power spectrum governed by the $\chi^2$-distribution with the number of degrees of freedom greater than two. In both the Fourier and EMD approaches, periods of $48.6 \pm 7.0$\,s (FUV) and $49.2 \pm 7.4$\,s (NUV) were found to be significant. The uncertainties for each mean modal period were estimated using the half-level width of a best-fit Gaussian function in the corresponding global wavelet spectra. Those uncertainties are connected with the instantaneous period drift in each EMD mode rather than instrumental effects. The non-periodic trends were excluded from this analysis. Thus, the flare periodicities observed in the FUV and NUV bands are clearly consistent between each other within the estimated uncertainties, which further strengthens the confidence that the detected QPP are intrinsic to the star and not a systematic error.

The statistical significance of the detected $\sim$50 second period is seen to be substantially higher in the NUV band. This is seen in both the Fourier and EMD analyses, which are intrinsically independent, thus ruling out the possibility that it is caused by the method. On the other hand, in the EMD power spectrum of the FUV signal, the first (shortest-period) mode near 20 seconds is seen at the 99\% confidence level too, while the corresponding Fourier peak is well below its level of significance. We note that the shortest time scale modes in the EMD analysis usually have abnormally distributed instantaneous amplitudes, hence their energies (sum of instantaneous amplitudes squared) do not follow the $\chi^2$-law \citep{2004RSPSA.460.1597W}. This makes the significance test performed in this study inapplicable to the shortest-period modes.  In addition, the distribution of the modal energies in the EMD spectrum may be corrupted by the so-called ``mode leakage'' problem, when limitations from the time resolution of the light curve impact the ability of EMD to properly differentiate between neighboring modal periods \citep{Rilling2009}.

Finally, the dither pattern of the observation during this flare induces a sinusoidal signal with a period of 120 seconds (a plot of the detector position over time for GJ 65 is available in a Python notebook associated with the online version of this paper).  Induced signals at, or aliased with, the dither pattern frequency are an important source of systematic error when using gPhoton data \citep{2018ApJS..238...25D}, especially near the detector edges or when at or near the local non-linearity threshold (which applies for this flare).  The strongest detected period at $\sim$50 seconds in both FUV and NUV bands is far enough away from any aliases of the dither pattern that it is highly unlikely these signals are induced by local non-linearity.  In fact, when conducting their wavelet analysis of GJ 65, \citet{2018MNRAS.475.2842D} found a 46 second period in the rise and late decay phases of the flare (where the count rate was below or barely above the local non-linearity threshold), and a ``weak $\sim$25 second period around flare maximum'', where the effects of depressed flux due to local non-linearity would be at their most severe.  While \citet{2018MNRAS.475.2842D} do not provide any uncertainty estimates for the detected periods, we note the 46 second period is a very good match to the period found in our analysis (and unlike \citet{2018MNRAS.475.2842D}, we analyzed both the FUV and NUV bands), while their 25 second period found only when count rate was highest may be the result of local non-linearity triggering off a 1:6 alias of the dither pattern of the observations, and could also match our second highest peak at $\sim$20 seconds.  All this leads us to disregard the shortest-period mode in Fig.\ \ref{emd_fuv} as an astrophysical source despite the formal statistical significance from the EMD analysis, and if anything, we suspect it may be caused by an induced signal due to local non-linearity during the peak of the flare that relates to the dither pattern of the telescope.

\subsubsection{QPP Physical Interpretation}
The period of a QPP is linked to a characteristic length scale and a velocity, for a given interpretation of the QPP mechanism (i.e. ``sausage'' or ``kink'' magnetohydrodynamic (MHD) modes). MHD waves behave differently at the flare loop tops and near the footpoints \citep{2016SSRv..200...75N}.  Standing, fast magnetoacoustic kink oscillations weakly perturb the loop density, while the perturbation of the loop velocity has nodes and antinodes near the footpoints and at the apex, respectively.  Fast sausage waves have stronger density perturbations higher up in the corona.  For compressive, slow magnetoacoustic waves near the loop footpoints, the perturbation of velocity is zero and the perturbation of density is maximized.

Provided the loop is large enough so that its apex is in the corona, kinks are not likely to modulate the chromospheric source. Sausage modes can modulate precipitation of accelerated electrons towards the footpoints by changing the magnetic mirror ratio in the loop via the ``Zaitsev-Stepanov'' mechanism \citep{2020STP.....6a...3K}.  For solar coronal loops the periods of sausage modes are usually 10-20 s. For standing slow modes (a variation of density near the footpoints), oscillation periods are usually longer (at least several minutes or more).  A third possible mechanism for QPP are periodically triggered magnetic reconnections in the corona. In this case, the coronal reconnection site is unstable to an external oscillation like a kink \citep{2006A&A...452..343N}, providing periodic precipitations of energetic particles down to the chromosphere and heating the plasma again.  The shortest value of kink periods observed in the solar corona are about 1 min \citep{2019ApJS..241...31N}, not too far off from the $\sim$50 second periods observed here.  Out of these three possible mechanisms, periodic triggering of reconnection by external MHD oscillation matches our observed QPP period best, but the others are not necessarily excluded, especially when considering potential differences in the QPP properties generated by these three mechanisms for M dwarfs compared to the Sun.

\section{Conclusion}\label{sec:conc}
The combination of high flare-to-star contrast in the UV, wide spatial coverage of GALEX, and fast temporal sampling of the photon events enabled by gPhoton make for a rich archive for flare science. The M5 dwarfs in the GJ 65 system are the brightest, active flare sources observed by GALEX for more than a few minutes. GJ 65 is thus the best opportunity in the GALEX data to detect and characterize low-energy M dwarf flares.  We have described 14 previously undetected flare events in the GALEX data, all with UV flare energies below $10^{29.5}$ ergs.  This is less than than the minimum-detectable flare energies by space-based surveys from Kepler, K2, and TESS, but our estimated FFD is broadly consistent with the predicted values based on extrapolated FFD from those larger ground- and space-based surveys.  We also time-resolve the largest detected flare from GJ 65, previously detected by \citet{2017AJ....154...67M} at the visit level.  Although the count rate exceeds the local non-linearity threshold during the flare, using the simultaneous FUV and NUV observations provided by GALEX, we detect strong, periodic variability during the flare itself that is not correlated with the dither pattern of the telescope.  Both the FUV and NUV bands show a strong signal with a period of $\sim$50 seconds, which is not seen in other nearby stars and is not correlated with known instrumental systematics.  We interpret this signal as a strong quasi-periodic pulsation associated with the flare.  Given the period of the QPP, the best matching mechanism based on typical timescales from solar observations would be a periodic triggering of reconnection caused by external MHD oscillation, since the shortest timescale of such kinks observed in the Sun's corona are $\sim$1 min and most closely match the $\sim$50 second period here.  However, the extent to which a flare mechanism can be identified solely by comparing timescales of Solar QPPs with observations of QPPs in M dwarfs is uncertain, so we consider this interpretation to be suggestive rather than definitive.

GALEX observations of tens of thousands of other cool stars, in addition to solar-like stars, are available in the gPhoton archive. Unlike GJ 65, most stars were not observed multiple times by GALEX, but the nearly complete sky coverage makes the GALEX archive a rich resource for UV flare science on a massive scale.  While GJ 65 might afford the best single source for low-energy flare studies using the gPhoton archive, the sheer number of stars with more than 10-minute long observations allows for ensemble investigations of flares across a wide range of stellar types at a time resolution of seconds, a sampling rate that is still inaccessible for the majority of targets being observed from space-based surveys.

\acknowledgments
We thank the anonymous referee whose comments improved the quality of this paper. This research is supported by NASA ADAP grant 80NSSC18K0084.  We thank Parke Loyd and Evgenya Shkolnik for discussions that greatly improved the quality of the paper.  This research is based on observations made with the Galaxy Evolution Explorer, obtained from the MAST data archive at the Space Telescope Science Institute, which is operated by the Association of Universities for Research in Astronomy, Inc., under NASA contract NAS 5–26555.  This research has made use of the SVO Filter Profile Service (\url{http://svo2.cab.inta-csic.es/theory/fps/}) supported from the Spanish MINECO through grant AYA2017-84089. DYK acknowledges support by the STFC consolidated grant ST/T000252/1 and the Ministry of Science and Higher Education of the Russian Federation.

\facility{GALEX}

\software{Astropy \citep{2013A&A...558A..33A,2018AJ....156..123A}, gPhoton \citep{2016ApJ...833..292M}}




\appendix
\section{\label{app_flareplots}Visit Light Curves and Detected Flares}
\begin{figure}
\includegraphics[width=0.8\textwidth]{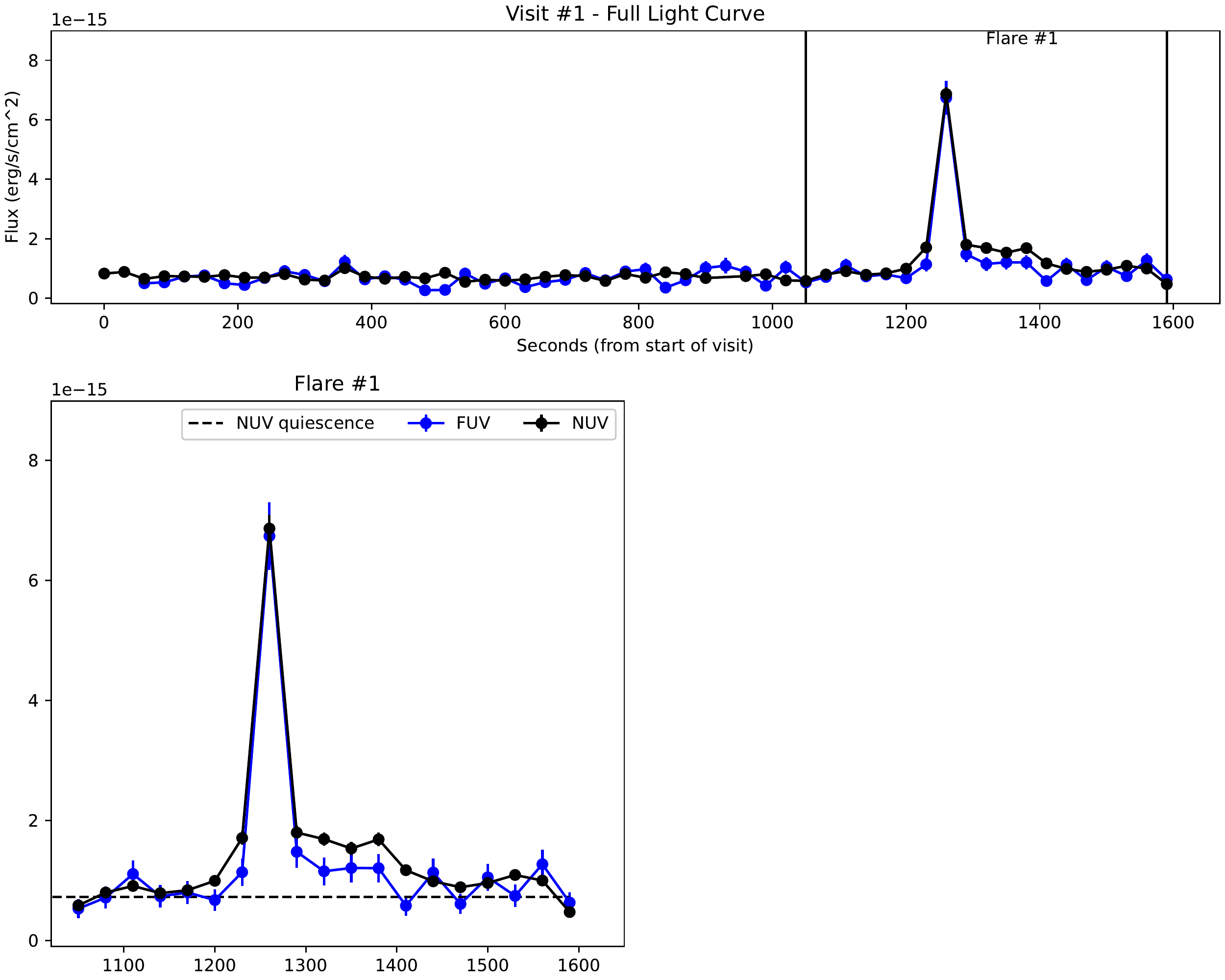}
\caption{Detected flares from Visit 1. Full visit (top), and zoomed plots centered on each detected flare event (bottom).\label{flarefig1}}
\end{figure}

\begin{figure}
\includegraphics[width=0.8\textwidth]{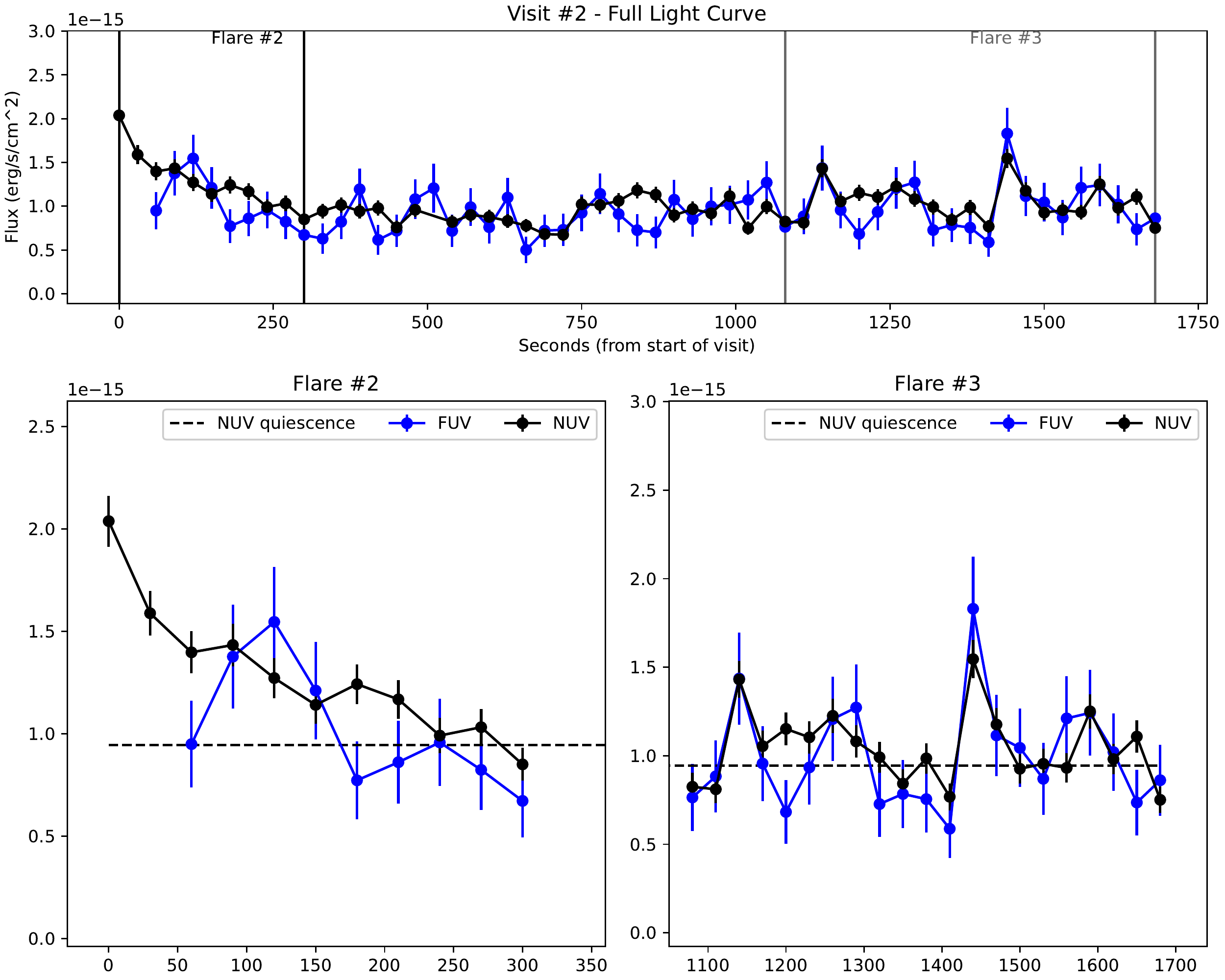}
\caption{Detected flares from Visit 2. Full visit (top), and zoomed plots centered on each detected flare event (bottom).\label{flarefig2}}
\end{figure}

\begin{figure}
\includegraphics[width=0.8\textwidth]{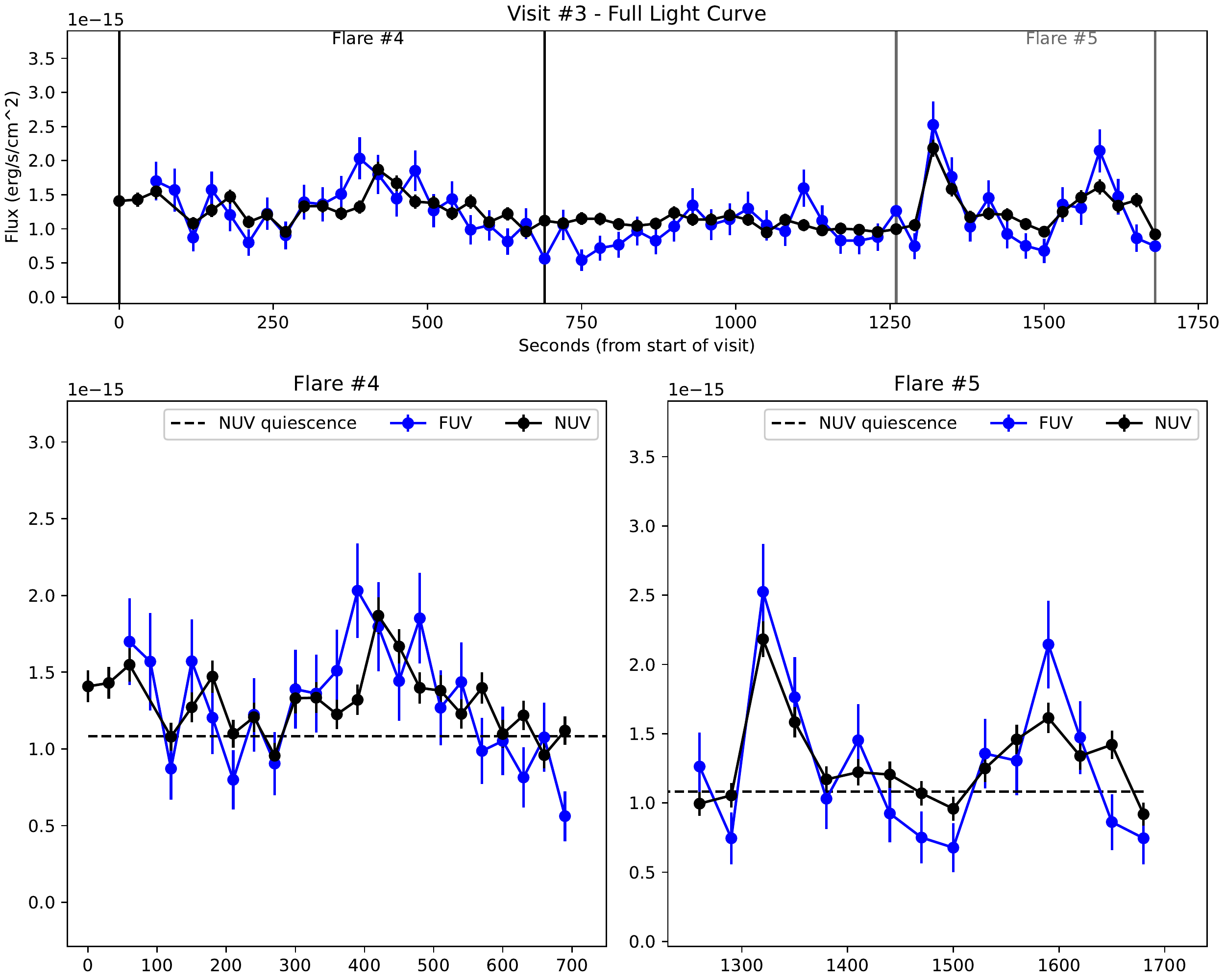}
\caption{Detected flares from Visit 3. Full visit (top), and zoomed plots centered on each detected flare event (bottom).\label{flarefig3}}
\end{figure}

\begin{figure}
\includegraphics[width=0.8\textwidth]{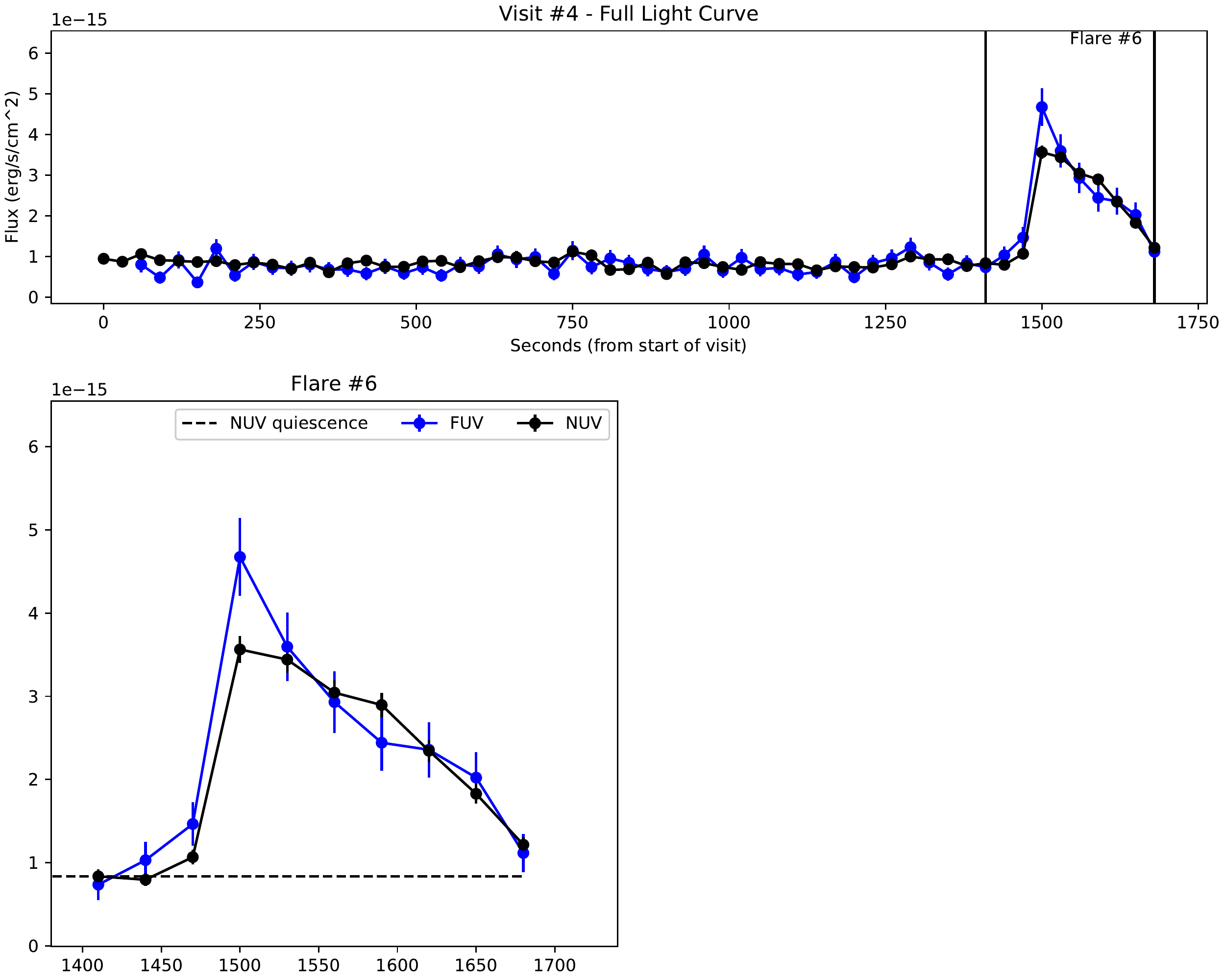}
\caption{Detected flares from Visit 4. Full visit (top), and zoomed plots centered on each detected flare event (bottom).\label{flarefig4}}
\end{figure}

\begin{figure}
\includegraphics[width=0.8\textwidth]{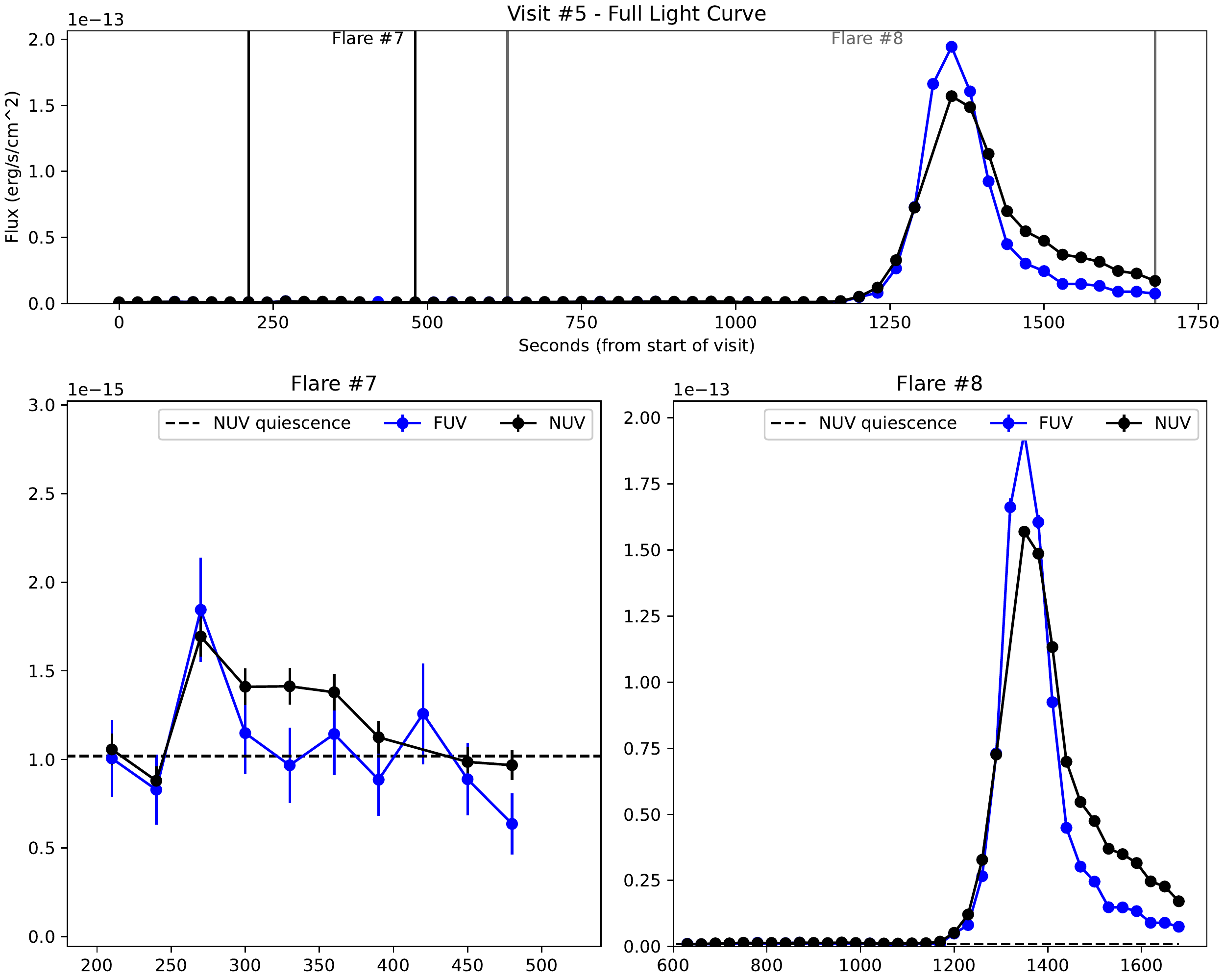}
\caption{Detected flares from Visit 5. Full visit (top), and zoomed plots centered on each detected flare event (bottom).  Flare 7 is the strongest flare detected in gPhoton for GJ 65. Although there is an apparent time evolution in the FUV-NUV flux ratio, the count rate exceeds the local non-linearity threshold by a significant margin, and thus any conclusions on FUV-NUV ratio cannot be made without applying a robust, accurate correction for missing fluxes in both bands (see Sec. \ref{flare8}). \label{flarefig5}}
\end{figure}

\begin{figure}
\includegraphics[width=0.8\textwidth]{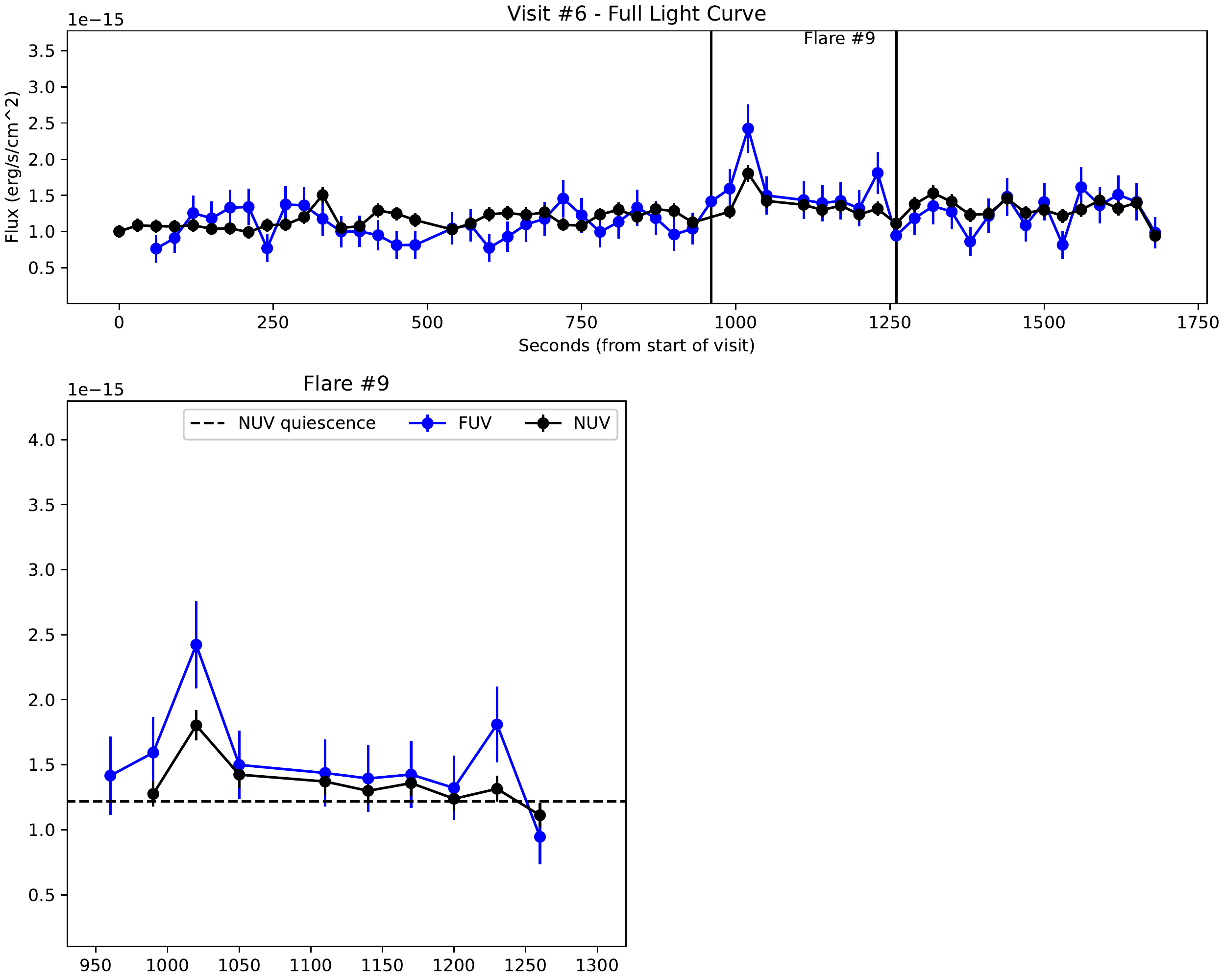}
\caption{Detected flares from Visit 6. Full visit (top), and zoomed plots centered on each detected flare event (bottom).\label{flarefig6}}
\end{figure}

\begin{figure}
\includegraphics[width=0.8\textwidth]{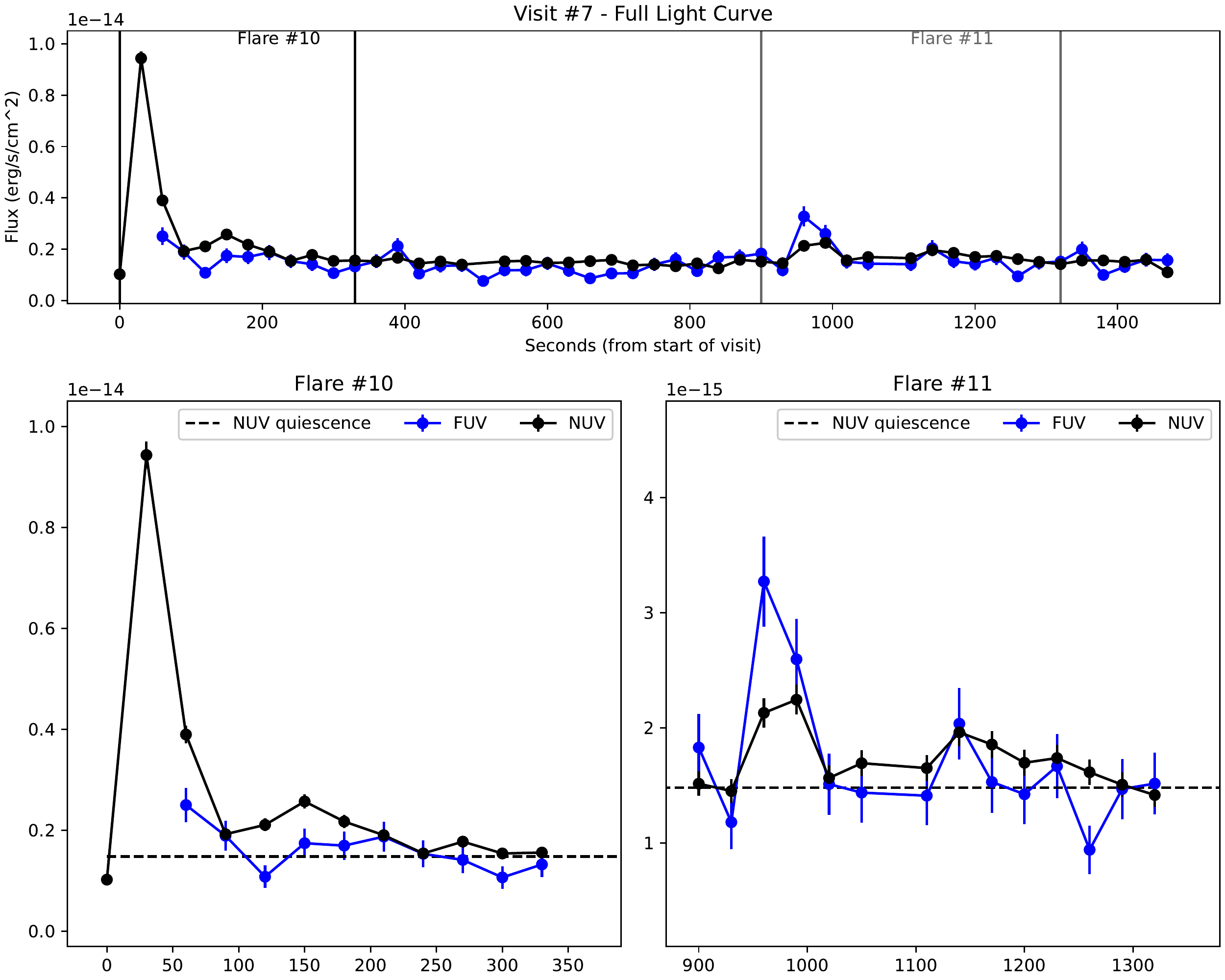}
\caption{Detected flares from Visit 7. Full visit (top), and zoomed plots centered on each detected flare event (bottom).  Note: Flare 10 is truncated so much that we do not include it in our analysis, but include it in Table \ref{flaretable} for completeness.  The apparent gap in the second flare event is caused by a time bin that falls below our threshold for effective exposure time coverage (that particular bin only has $\sim16$ seconds of effective exposure time within the 30-second bin).\label{flarefig7}}
\end{figure}

\begin{figure}
\includegraphics[width=0.8\textwidth]{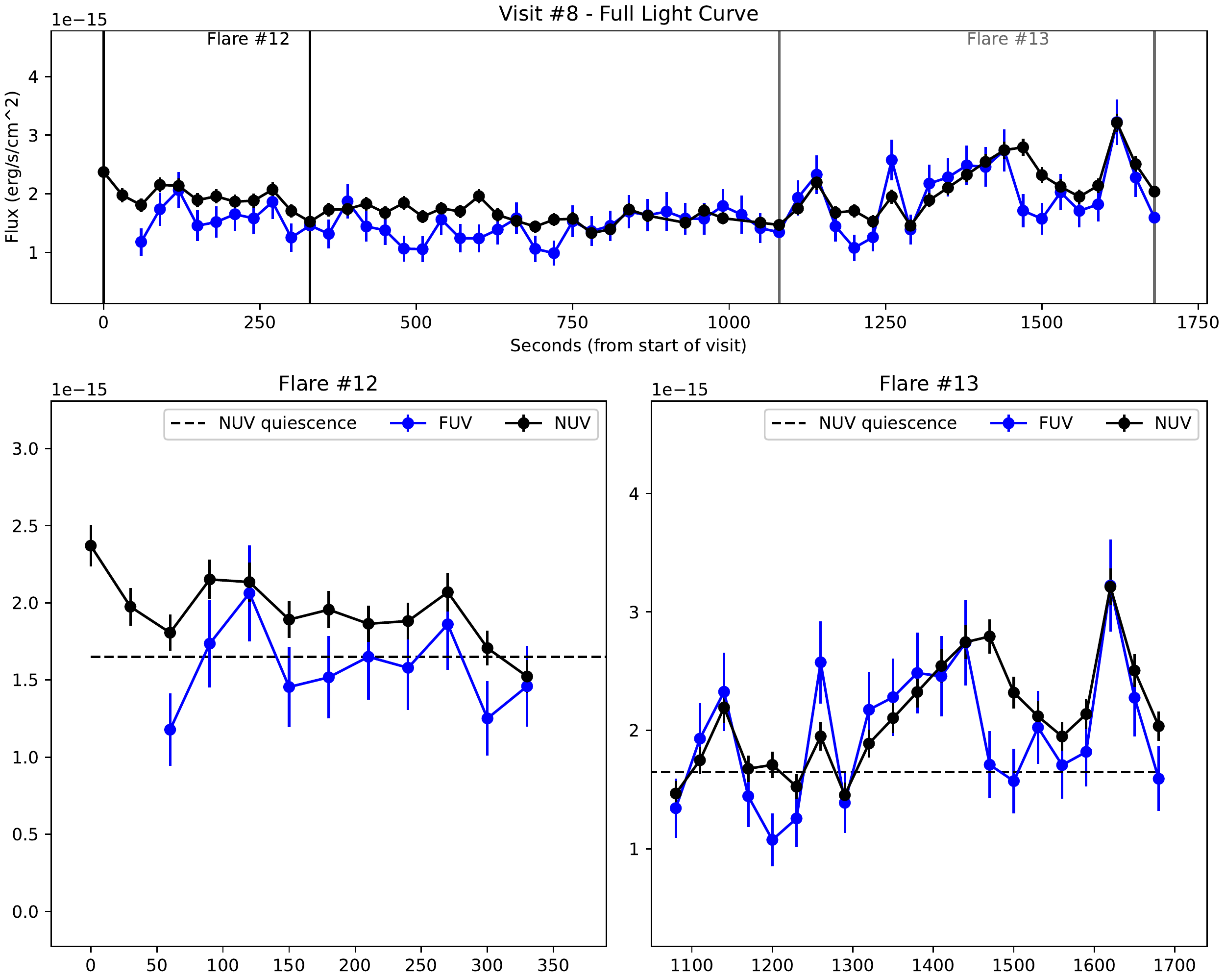}
\caption{Detected flares from Visit 8. Full visit (top), and zoomed plots centered on each detected flare event (bottom).\label{flarefig8}}
\end{figure}

\begin{figure}
\includegraphics[width=0.8\textwidth]{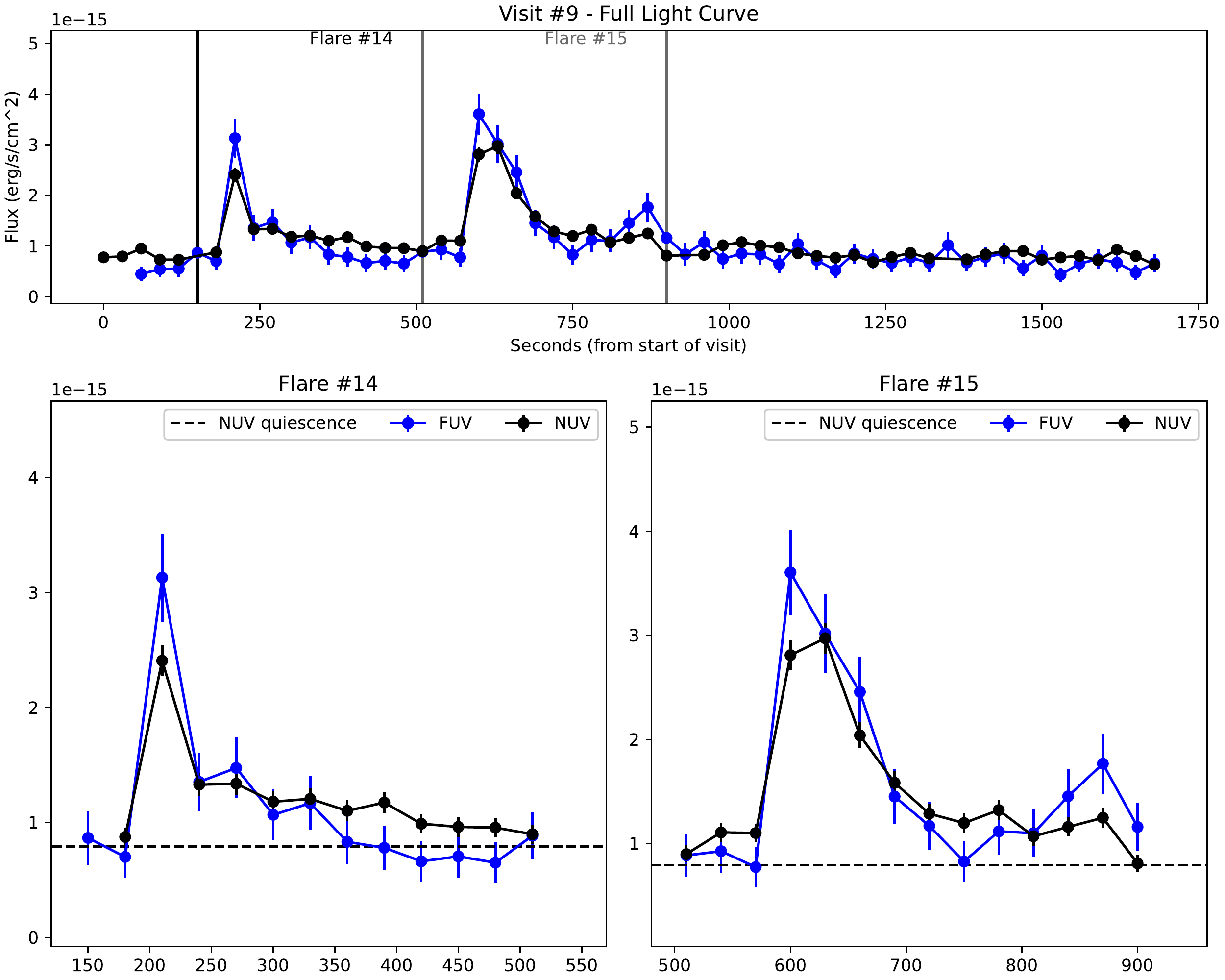}
\caption{Detected flares from Visit 9. Full visit (top), and zoomed plots centered on each detected flare event (bottom).\label{flarefig9}}
\end{figure}

Figures \ref{flarefig1}, \ref{flarefig2}, \ref{flarefig3}, \ref{flarefig4}, \ref{flarefig5}, \ref{flarefig6}, \ref{flarefig7}, \ref{flarefig8} and \ref{flarefig9} plot the detected flares from the GJ 65 visits we analyzed.  The top panels show the full visit light curves from both the FUV and NUV bands.  The bottom panels show a zoom-in on the flare ranges themselves (each visit contains one or two flare events).  The INFF value in the NUV is shown for reference.  In all cases, the FUV INFF values are very similar to the NUV value.

\begin{thebibliography}{}

\bibitem[Astropy Collaboration et al.(2013)]{2013A&A...558A..33A} Astropy Collaboration, Robitaille, T.~P., Tollerud, E.~J., et al.\ 2013, \aap, 558, A33
\bibitem[Astropy Collaboration et al.(2018)]{2018AJ....156..123A} Astropy Collaboration, Price-Whelan, A.~M., Sip{\H o}cz, B.~M., et al.\ 2018, \aj, 156, 123
\bibitem[Atri(2017)]{2017MNRAS.465L..34A} Atri, D.\ 2017, \mnras, 465, L34
\bibitem[Avrett, \& Loeser(2008)]{2008ApJS..175..229A} Avrett, E.~H., \& Loeser, R.\ 2008, \apjs, 175, 229
\bibitem[Bailer-Jones et al.(2018)]{2018AJ....156...58B} Bailer-Jones, C.~A.~L., Rybizki, J., Fouesneau, M., et al.\ 2018, \aj, 156, 58
\bibitem[Barnes et al.(2017)]{2017MNRAS.471..811B} Barnes, J.~R., Jeffers, S.~V., Haswell, C.~A., et al.\ 2017, \mnras, 471, 811
\bibitem[Beskin et al.(2017)]{2017PASA...34...10B} Beskin, G., Karpov, S., Plokhotnichenko, V., et al.\ 2017, \pasa, 34, e010
\bibitem[Bopp \& Moffett(1973)]{1973ApJ...185..239B} Bopp, B.~W., \& Moffett, T.~J.\ 1973, \apj, 185, 239
\bibitem[Boudreaux et al.(2017)]{2017ApJ...845..171B} Boudreaux, T.~M., Barlow, B.~N., Fleming, S.~W., et al.\ 2017, \apj, 845, 171
\bibitem[Brasseur et al.(2019)]{brasseur2019} Brasseur, C.~E., Osten, R.~A., \& Fleming, S.~W.\ 2019, \apj, 883, 88
\bibitem[Broomhall et al.(2019)]{2019ApJS..244...44B} Broomhall, A.-M., Davenport, J.~R.~A., Hayes, L.~A., et al.\ 2019, \apjs, 244, 44
\bibitem[Chadney et al.(2017)]{2017A&A...608A..75C} Chadney, J.~M., Koskinen, T.~T., Galand, M., et al.\ 2017, \aap, 608, A75
\bibitem[Chang et al.(2020)]{2020MNRAS.491...39C} Chang, S.-W., Wolf, C., \& Onken, C.~A.\ 2020, \mnras, 491, 39
\bibitem[Davenport et al.(2012)]{2012ApJ...748...58D} Davenport, J.~R.~A., Becker, A.~C., Kowalski, A.~F., et al.\ 2012, \apj, 748, 58
\bibitem[Davenport et al.(2014)]{2014ApJ...797..122D} Davenport, J.~R.~A., Hawley, S.~L., Hebb, L., et al.\ 2014, \apj, 797, 122
\bibitem[Davenport(2016)]{2016ApJ...829...23D} Davenport, J.~R.~A.\ 2016, \apj, 829, 23 
\bibitem[de la Vega, \& Bianchi(2018)]{2018ApJS..238...25D} de la Vega, A., \& Bianchi, L.\ 2018, \apjs, 238, 25
\bibitem[Dere et al.(2009)]{2009A&A...498..915D} Dere, K.~P., Landi, E., Young, P.~R., et al.\ 2009, \aap, 498, 915
\bibitem[Dong et al.(2018)]{2018PNAS..115..260D} Dong, C., Jin, M., Lingam, M., et al.\ 2018, Proceedings of the National Academy of Science, 115, 260
\bibitem[Doyle et al.(2018)]{2018MNRAS.475.2842D} Doyle, J.~G., Shetye, J., Antonova, A.~E., et al.\ 2018, \mnras, 475, 2842
\bibitem[Doyle et al.(2019)]{2019MNRAS.tmp.2115D} Doyle, L., Ramsay, G., Doyle, J.~G., et al.\ 2019, \mnras, 2115
\bibitem[Ebeling(2003)]{2003MNRAS.340.1269E} Ebeling, H.\ 2003, \mnras, 340, 1269
\bibitem[Fontenla et al.(2016)]{2016ApJ...830..154F} Fontenla, J.~M., Linsky, J.~L., Witbrod, J., et al.\ 2016, \apj, 830, 154
\bibitem[Frasca et al.(2016)]{2016A&A...594A..39F} Frasca, A., Molenda-{\.Z}akowicz, J., De Cat, P., et al.\ 2016, \aap, 594, A39
\bibitem[Froning et al.(2019)]{2019ApJ...871L..26F} Froning, C.~S., Kowalski, A., France, K., et al.\ 2019, \apjl, 871, L26
\bibitem[Gaia Collaboration et al.(2018)]{2018A&A...616A...1G} Gaia Collaboration, Brown, A.~G.~A., Vallenari, A., et al.\ 2018, \aap, 616, A1
\bibitem[Gehrels(1986)]{1986ApJ...303..336G} Gehrels, N.\ 1986, \apj, 303, 336
\bibitem[G{\"u}nther et al.(2020)]{2020AJ....159...60G} G{\"u}nther, M.~N., Zhan, Z., Seager, S., et al.\ 2020, \aj, 159, 60
\bibitem[Gizis et al.(2017)]{2017ApJ...845...33G} Gizis, J.~E., Paudel, R.~R., Mullan, D., et al.\ 2017, \apj, 845, 33
\bibitem[Haisch et al.(1991)]{1991ARA&A..29..275H} Haisch, B., Strong, K.~T., \& Rodono, M.\ 1991, \araa, 29, 275
\bibitem[Hannah et al.(2011)]{2011SSRv..159..263H} Hannah, I.~G., Hudson, H.~S., Battaglia, M., et al.\ 2011, \ssr, 159, 263 
\bibitem[Hawley et al.(2003)]{2003ApJ...597..535H} Hawley, S.~L., Allred, J.~C., Johns-Krull, C.~M., et al.\ 2003, \apj, 597, 535
\bibitem[Hawley et al.(2014)]{2014ApJ...797..121H} Hawley, S.~L., Davenport, J.~R.~A., Kowalski, A.~F., et al.\ 2014, \apj, 797, 121 
\bibitem[Hilton(2011)]{2011PhDT.......144H} Hilton, E.~J.\ 2011, Ph.D.~Thesis
\bibitem[Howard et al.(2019)]{2019ApJ...881....9H} Howard, W.~S., Corbett, H., Law, N.~M., et al.\ 2019, \apj, 881, 9
\bibitem[Howard et al.(2021)]{howard2021} Howard, W.~S., MacGregor, M.~A.\ 2021, arXiv preprint arXiv:2110.13155
\bibitem[Huang et al.(1998)]{1998RSPSA.454..903H} Huang, N.~E., Shen, Z., Long, S.~R., et al.\ 1998, Proceedings of the Royal Society of London Series A, 454, 903
\bibitem[Huang \& Wu(2008)]{2008RvGeo..46.2006H} Huang, N.~E., \& Wu, Z.\ 2008, Reviews of Geophysics, 46, RG2006
\bibitem[Jackman et al.(2019)]{2019MNRAS.482.5553J} Jackman, J.~A.~G., Wheatley, P.~J., Pugh, C.~E., et al.\ 2019, \mnras, 482, 5553
\bibitem[Kaltenegger(2017)]{2017ARA&A..55..433K} Kaltenegger, L.\ 2017, \araa, 55, 433
\bibitem[Kervella et al.(2016)]{2016A&A...593A.127K} Kervella, P., M{\'e}rand, A., Ledoux, C., Demory, B.-O., \& Le Bouquin, J.-B.\ 2016, \aap, 593, A127
\bibitem[Kochukhov, \& Lavail(2017)]{2017ApJ...835L...4K} Kochukhov, O., \& Lavail, A.\ 2017, \apj, 835, L4
\bibitem[Kolotkov et al.(2015)]{2015A&A...574A..53K} Kolotkov, D.~Y., Nakariakov, V.~M., Kupriyanova, E.~G., et al.\ 2015, \aap, 574, A53
\bibitem[Kolotkov et al.(2016)]{2016A&A...592A.153K} Kolotkov, D.~Y., Anfinogentov, S.~A., \& Nakariakov, V.~M.\ 2016, \aap, 592, A153
\bibitem[Kolotkov et al.(2018)]{2018ApJ...858L...3K} Kolotkov, D.~Y., Pugh, C.~E., Broomhall, A.-M., et al.\ 2018, \apjl, 858, L3
\bibitem[Kowalski et al.(2012)]{2012SoPh..277...21K} Kowalski, A.~F., Hawley, S.~L., Holtzman, J.~A., et al.\ 2012, \solphys, 277, 21
\bibitem[Kowalski et al.(2013)]{2013ApJS..207...15K} Kowalski, A.~F., Hawley, S.~L., Wisniewski, J.~P., et al.\ 2013, \apjs, 207, 15
\bibitem[Kupriyanova et al.(2010)]{2010SoPh..267..329K} Kupriyanova, E.~G., Melnikov, V.~F., Nakariakov, V.~M., et al.\ 2010, \solphys, 267, 329
\bibitem[Kupriyanova et al.(2020)]{2020STP.....6a...3K} Kupriyanova, E., Kolotkov, D., Nakariakov, V., et al.\ 2020, Solar-Terrestrial Physics, 6, 3
\bibitem[Kuznetsov \& Kolotkov(2021)]{2021ApJ...912...81K} Kuznetsov, A.~A. \& Kolotkov, D.~Y.\ 2021, \apj, 912, 81. doi:10.3847/1538-4357/abf569
\bibitem[Lammer et al.(2007)]{2007AsBio...7..185L} Lammer, H., Lichtenegger, H.~I.~M., Kulikov, Y.~N., et al.\ 2007, Astrobiology, 7, 185 
\bibitem[Linsky et al.(2013)]{2013ApJ...766...69L} Linsky, J.~L., France, K., \& Ayres, T.\ 2013, \apj, 766, 69
\bibitem[Loyd et al.(2018a)]{2018ApJ...867...71L} Loyd, R.~O.~P., France, K., Youngblood, A., et al.\ 2018, \apj, 867, 71
\bibitem[Loyd et al.(2018b)]{2018ApJ...867...70L} Loyd, R.~O.~P., Shkolnik, E.~L., Schneider, A.~C., et al.\ 2018, \apj, 867, 70
\bibitem[MacGregor et al.(2021)]{2021ApJ...911L..25M} MacGregor, M.~A., Weinberger, A.~J., Loyd, R.~O.~P., et al.\ 2021, \apjl, 911, L25. doi:10.3847/2041-8213/abf14c
\bibitem[Maehara et al.(2015)]{2015EP&S...67...59M} Maehara, H., Shibayama, T., Notsu, Y., et al.\ 2015, Earth, Planets, and Space, 67, 59
\bibitem[Mann et al.(2019)]{2019ApJ...871...63M} Mann, A.~W., Dupuy, T., Kraus, A.~L., et al.\ 2019, \apj, 871, 63
\bibitem[Mathioudakis et al.(2003)]{2003A&A...403.1101M} Mathioudakis, M., Seiradakis, J.~H., Williams, D.~R., et al.\ 2003, \aap, 403, 1101
\bibitem[Martin et al.(2005)]{2005ApJ...619L...1M} Martin, D.~C., Fanson, J., Schiminovich, D., et al.\ 2005, \apjl, 619, L1
\bibitem[Miles \& Shkolnik(2017)]{2017AJ....154...67M} Miles, B.~E., \& Shkolnik, E.~L.\ 2017, \aj, 154, 67
\bibitem[Million et al.(2016)]{2016ApJ...833..292M} Million, C., Fleming, S.~W., Shiao, B., et al.\ 2016, \apj, 833, 292
\bibitem[Morrissey et al.(2007)]{2007ApJS..173..682M} Morrissey, P., Conrow, T., Barlow, T.~A., et al.\ 2007, \apjs, 173, 682
\bibitem[Nakariakov et al.(2006)]{2006A&A...452..343N} Nakariakov, V.~M., Foullon, C., Verwichte, E., et al.\ 2006, \aap, 452, 343
\bibitem[Nakariakov et al.(2016)]{2016SSRv..200...75N} Nakariakov, V.~M., Pilipenko, V., Heilig, B., et al.\ 2016, \ssr, 200, 75
\bibitem[Namekata et al.(2017)]{2017ApJ...851...91N} Namekata, K., Sakaue, T., Watanabe, K., et al.\ 2017, \apj, 851, 91
\bibitem[Nechaeva et al.(2019)]{2019ApJS..241...31N} Nechaeva, A., Zimovets, I.~V., Nakariakov, V.~M., et al.\ 2019, \apjs, 241, 31
\bibitem[O'Malley-James, \& Kaltenegger(2017)]{2017MNRAS.469L..26O} O'Malley-James, J.~T., \& Kaltenegger, L.\ 2017, \mnras, 469, L26
\bibitem[Osten, \& Brown(1999)]{1999ApJ...515..746O} Osten, R.~A., \& Brown, A.\ 1999, \apj, 515, 746
\bibitem[Osten et al.(2012)]{2012ApJ...754....4O} Osten, R.~A., Kowalski, A., Sahu, K., et al.\ 2012, \apj, 754, 4
\bibitem[Osten \& Wolk(2015)]{2015ApJ...809...79O} Osten, R.~A. \& Wolk, S.~J.\ 2015, \apj, 809, 79. doi:10.1088/0004-637X/809/1/79
\bibitem[Panagi, \& Andrews(1995)]{1995MNRAS.277..423P} Panagi, P.~M., \& Andrews, A.~D.\ 1995, \mnras, 277, 423
\bibitem[Parks \& Winckler(1969)]{1969ApJ...155L.117P} Parks, G.~K., \& Winckler, J.~R.\ 1969, \apjl, 155, L117
\bibitem[Pazzani, \& Rodono(1981)]{1981Ap&SS..77..347P} Pazzani, V., \& Rodono, M.\ 1981, \apss, 77, 347
\bibitem[Peacock et al.(2019)]{2019ApJ...871..235P} Peacock, S., Barman, T., Shkolnik, E.~L., et al.\ 2019, \apj, 871, 235
\bibitem[Pearce, \& Harrison(1990)]{1990A&A...228..513P} Pearce, G., \& Harrison, R.~A.\ 1990, \aap, 228, 513
\bibitem[Pugh et al.(2015)]{2015ApJ...813L...5P} Pugh, C.~E., Nakariakov, V.~M., \& Broomhall, A.-M.\ 2015, \apjl, 813, L5
\bibitem[Pugh et al.(2016)]{2016MNRAS.459.3659P} Pugh, C.~E., Armstrong, D.~J., Nakariakov, V.~M., et al.\ 2016, Monthly Notices of the Royal Astronomical Society, 459, 3659
\bibitem[Pugh et al.(2017)]{2017A&A...602A..47P} Pugh, C.~E., Broomhall, A.-M., \& Nakariakov, V.~M.\ 2017, \aap, 602, A47
\bibitem[Raetz et al.(2020)]{2020A&A...637A..22R} Raetz, S., Stelzer, B., Damasso, M., et al.\ 2020, \aap, 637, A22
\bibitem[Ranjan \& Sasselov(2016)]{2016AsBio..16...68R} Ranjan, S., \& Sasselov, D.~D.\ 2016, Astrobiology, 16, 68
\bibitem[Ranjan et al.(2017)]{2017ApJ...843..110R} Ranjan, S., Wordsworth, R., \& Sasselov, D.~D.\ 2017, \apj, 843, 110
\bibitem[Rilling \& Flandrin(2009)]{Rilling2009}Rilling, Gabriel \& Flandrin, Patrick\ 2009, Advances in Adaptive Data Analysis, 1, 43-59
\bibitem[Ribas et al.(2016)]{2016A&A...596A.111R} Ribas, I., Bolmont, E., Selsis, F., et al.\ 2016, \aap, 596, A111
\bibitem[Robinson et al.(2005)]{2005ApJ...633..447R} Robinson, R.~D., Wheatley, J.~M., Welsh, B.~Y., et al.\ 2005, \apj, 633, 447
\bibitem[Rodono(1974)]{1974A&A....32..337R} Rodono, M.\ 1974, \aap, 32, 337
\bibitem[The SVO Filter Profile Service(2020)]{svoprof} The SVO Filter Profile Service, Rodrigo, C., Solano, E., Bayo, A. http://ivoa.net/documents/Notes/SVOFPS/index.html
\bibitem[Schmitt et al.(2016)]{2016A&A...589A..48S} Schmitt, J.~H.~M.~M., Kanbach, G., Rau, A., \& Steinle, H.\ 2016, \aap, 589, A48
\bibitem[Schrijver et al.(2012)]{2012JGRA..117.8103S} Schrijver, C.~J., Beer, J., Baltensperger, U., et al.\ 2012, Journal of Geophysical Research (Space Physics), 117, A08103
\bibitem[Segura et al.(2010)]{2010AsBio..10..751S} Segura, A., Walkowicz, L.~M., Meadows, V., Kasting, J., \& Hawley, S.\ 2010, Astrobiology, 10, 751
\bibitem[Sim{\~o}es et al.(2015)]{2015SoPh..290.3625S} Sim{\~o}es, P.~J.~A., Hudson, H.~S., \& Fletcher, L.\ 2015, \solphys, 290, 3625
\bibitem[Tarter et al.(2007)]{2007AsBio...7...30T} Tarter, J.~C., Backus, P.~R., Mancinelli, R.~L., et al.\ 2007, Astrobiology, 7, 30
\bibitem[Tilley et al.(2019)]{2019AsBio..19...64T} Tilley, M.~A., Segura, A., Meadows, V., et al.\ 2019, Astrobiology, 19, 64
\bibitem[Trainer et al.(2006)]{2006PNAS..10318035T} Trainer, M.~G., Pavlov, A.~A., Dewitt, H.~L., et al.\ 2006, Proceedings of the National Academy of Science, 103, 18035
\bibitem[van den Oord et al.(1996)]{1996A&A...310..908V} van den Oord, G.~H.~J., Doyle, J.~G., Rodono, M., et al.\ 1996, \aap, 310, 908
\bibitem[Vaughan(2005)]{2005A&A...431..391V} Vaughan, S.\ 2005, \aap, 431, 391
\bibitem[Venot et al.(2016)]{2016ApJ...830...77V} Venot, O., Rocchetto, M., Carl, S., Roshni Hashim, A., \& Decin, L.\ 2016, \apj, 830, 77
\bibitem[Walkowicz et al.(2011)]{2011AJ....141...50W} Walkowicz, L.~M., Basri, G., Batalha, N., et al.\ 2011, \aj, 141, 50
\bibitem[Welsh et al.(2005)]{2005AJ....130..825W} Welsh, B.~Y., Wheatley, J.~M., Heafield, K., et al.\ 2005, \aj, 130, 825
\bibitem[Welsh et al.(2006)]{2006A&A...458..921W} Welsh, B.~Y., Wheatley, J., Browne, S.~E., et al.\ 2006, \aap, 458, 921
\bibitem[Welsh et al.(2007)]{2007ApJS..173..673W} Welsh, B.~Y., Wheatley, J.~M., Seibert, M., et al.\ 2007, \apjs, 173, 673
\bibitem[Yang et al.(2018)]{2018ApJ...859...87Y} Yang, H., Liu, J., Qiao, E., et al.\ 2018, \apj, 859, 87
\bibitem[Wu \& Huang(2004)]{2004RSPSA.460.1597W} Wu, Z., \& Huang, N.~E.\ 2004, Proceedings of the Royal Society of London Series A, 460, 1597
\bibitem[Yamashiki et al.(2019)]{2019ApJ...881..114Y} Yamashiki, Y.~A., Maehara, H., Airapetian, V., et al.\ 2019, \apj, 881, 114
\bibitem[Yang et al.(2017)]{2017ApJ...849...36Y} Yang, H., Liu, J., Gao, Q., et al.\ 2017, \apj, 849, 36
\end{thebibliography}
\end{document}